\documentclass[10pt,letterpaper]{article}
\usepackage[top=0.85in,left=1in,footskip=0.75in]{geometry}

\usepackage{mathtools}
\usepackage{changepage}

\usepackage[utf8x]{inputenc}

\usepackage{textcomp,marvosym}

\usepackage{fixltx2e}

\usepackage{amssymb, amsmath}

\usepackage{cite}

\usepackage{nameref,hyperref}

\usepackage{multirow}

\usepackage{microtype}
\DisableLigatures[f]{encoding = *, family = * }

\raggedright
\usepackage[aboveskip=1pt,labelfont=bf,labelsep=period,justification=raggedright,singlelinecheck=off]{caption}

\makeatletter
\renewcommand{\@biblabel}[1]{\quad#1.}
\makeatother

\date{}

\usepackage{lastpage,fancyhdr,graphicx}
\graphicspath{{./images/}}
\usepackage{epstopdf}
\fancyheadoffset[L]{2.25in}
\fancyfootoffset[L]{2.25in}
\DeclarePairedDelimiter\floor{\lfloor}{\rfloor}

\begin{document}
\vspace*{0.2in}

\begin{flushleft}
{\Large
\textbf\newline{Cortical Circuits from Scratch: A Metaplastic Architecture for the Emergence of Lognormal Firing Rates and Realistic Topology} % Please use "title case" (capitalize all terms in the title except conjunctions, prepositions, and articles).
}
\newline
\\
Zo{\"e} Tosi\textsuperscript{1*},
John Beggs\textsuperscript{1,2}
\\
\bigskip
\textbf{1} Dept. Cognitive Science, Indiana University, Bloomington, IN, USA
\\
\textbf{2} Dept. Physics, Indiana University, Bloomington, IN, USA
\\
\bigskip

% Insert additional author notes using the symbols described below. Insert symbol callouts after author names as necessary.
% 
% Remove or comment out the author notes below if they aren't used.

% Use the asterisk to denote corresponding authorship and provide email address in note below.
* ztosi@iu.edu

\end{flushleft}
\section*{Abstract}

Our current understanding of neuroplasticity paints a picture of a complex interconnected system of dependent processes which shape cortical structure so as to produce an efficient information processing system. Indeed, the cooperation of these processes is associated with robust, stable, adaptable networks with characteristic features of activity and synaptic topology. However, combining the actions of these mechanisms in models has proven exceptionally difficult and to date no model has been able to do so without significant hand-tuning. Until such a model exists that can successfully combine these mechanisms to form a stable circuit with realistic features, our ability to study neuroplasticity in the context of (more realistic) dynamic networks and potentially reap whatever rewards these features and mechanisms imbue biological networks with is hindered. We introduce a model which combines five known plasticity mechanisms that act on the network as well as a unique \emph{metaplastic} mechanism which acts on other plasticity mechanisms, to produce a neural circuit model which is both stable and capable of broadly reproducing many characteristic features of cortical networks. The MANA (metaplastic artificial neural architecture) represents the first model of its kind in that it is able to self-organize realistic, nonrandom features of cortical networks, from a null initial state (no synaptic connectivity  or neuronal differentiation) with no hand-tuning of relevant variables. In the same vein as models like the SORN (self-organizing recurrent network) MANA represents further progress toward the reverse engineering of the brain at the network level.

% Please keep the Author Summary between 150 and 200 words
% Use first person. PLOS ONE authors please skip this step. 
% Author Summary not valid for PLOS ONE submissions.   
\section*{Author Summary}
Neural circuits are known to possess specific nonrandom features of wiring and firing behavior across brain areas and species, and though a complete picture is out of reach significant amounts of information have been uncovered. Furthermore, a clearer picture of the known mechanisms ostensibly responsible for those features is emerging. It is thought that the nature of these features and mechanisms underlie the exceptional and efficient computational abilities of neural circuits. We introduce an architecture which self-organizes a wide array of known circuit features and complex nonrandom topology from a null initial state (no recurrent synaptic connections; uniform target firing rates), using known plasticity mechanisms where possible and introducing new ones where needed. In particular we introduce a metaplastic rule for self-organizing lognormally distributed target firing rates. In order to harness the possible benefits conferred by the features and self-organizing/adaptive mechanisms of neuroplasticity, a stable network capable of manifesting those features solely through mechanism is a prerequisite. We introduce just such a network in the form of MANA (Metaplastic Artificial Neural Architecture). 

%\linenumbers

% Use "Eq" instead of "Equation" for equation citations.
\section*{Introduction}

\subsection*{Motivation and Goals}
\label{MotsAndGoals}

What makes brains especially powerful, efficient and capable information processors? What about them so easily enables dynamic, real-time learning and cognition? In his 2007 paper \emph{What can AI get from Neuroscience?}, Steve Potter compares modern artificial intelligence to a hypothetical group of energy researchers who are aware of an alien power-plant discovered in the jungle which appears to provide virtually limitless clean power (the brain in this analogy), but who largely ignore it in favor of more tried and true techniques \cite{potter2007can}. Since 2007, convolutional neural networks, which vaguely mimic the staged feed-forward aspects of processing in visual cortex, have come to dominate computer vision and related domains of artificial intelligence due to their profound success \cite{krizhevsky2012imagenet}\cite{ILSVRC2016}. However, deep learning, though impressive, is a crude approximation of its biological inspiration. The question then remains: What other aspects of biological neural networks--if successfully reverse engineered--might lead to the next revolution? Despite the massive success of deep learning, AI researchers have largely avoided further attempts at reverse engineering the genuine article in a systematic way. 
 
 Living neural circuits have a very particular set of qualities and properties which characterize them including: lognormal firing rate \cite{hromadka2008sparse, mizuseki2013preconfigured, buzsaki2014log, nigam2016} and excitatory (inhibitory)synaptic weight distributions \cite{song2005highly, lefort2009excitatory, feldmeyer2002synaptic} (\cite{borst1994large, brussaard1997plasticity, nusser1997differences}), the over-representation of tightly connected clusters\cite{perin2011synaptic} and particular triadic motifs\cite{song2005highly},  the high in-degree and reduced inhibition of highly active neurons\cite{yassin2010embedded}\cite{benedetti2012differential}, and functional specialization conforming to certain organizational motifs \cite{harris2015neocortical}. If any of these qualities confer benefits to the processing or retention of information, then it is reasonable to assume that reproducing them (and the processes which lead to them) may confer those benefits in an artificial model. Notably, the processes which cause these features to arise are of equal importance since the degree to which many of these features are themselves beneficial or merely a side effect of mechanisms which are beneficial is unclear. In either case having a model which reproduces a wide array of circuit features and which does so through mechanism is an excellent starting point for the assessment of how those mechanisms/features contribute to the functioning of a circuit and the degree to which they may be useful if adapted for artificial intelligence purposes. Conforming to these constraints and demanding the reproduction of so many features, however, is a daunting task and amounts to reverse engineering cortex at the network level. Here we attempt to engage in that task by designing an artificial spiking neural network model which is able to manifest all of the aforementioned features from a completely null starting state and without significant direct hand tuning of relevant features. 
\paragraph{}
MANA uses 5 known mechanisms of plasticity and one hypothetical metaplastic mechanism--referred to as such since it is responsible for dynamically governing the evolution of the homeostatic set-points of other plasticity mechanisms (i.e. it is a plasticity mechanism of plasticity mechanisms as opposed to an agent of plastic change acting directly on network properties) and has no direct \emph{known} analog in living brains. This combination allows a circuit which replicates a wide variety of features to be self-organized from a null initial state whereby even target firing rates for each cell are not significantly specified prior to simulation. Specifically MANA is initialized with \emph{no synaptic connections and uniform target firing rates (TFRs) amongst its neurons}, such that the resulting synaptic topology and firing rate distribution are completely the result of plasticity driven growth and pruning, the metaplastic mechanism and the synergy of all plasticity mechanisms involved. We focus here \emph{only on the attaining of a great many different features from mechanism and not the computational aspects of the circuit} as prior to this work no model existed which could manifest the number of circuit features specified here through mechanism alone. Merely creating a model which does is itself a major undertaking, and the entire focus of this paper. Before the computational power of such a circuit can be tested, before certain mechanisms or features can be deemed superfluous, before any further investigation with respect to how the synergy of different mechanisms combine to manifest certain features, a model which \emph{can} manifest them through mechanism alone \emph{must first exist}. Detailing the first of that class of models is the subject of this paper.

\subsection*{Context and Other Work}

Crucial to the development and self-organization of any neural circuit is the differentiation of neurons and synapses into distinct functional roles. Differences in connectivity patterns and cortical cell classes improve information encoding by broadening the available strategies for information processing \cite{harris2013cortical}, while simultaneously similar motifs in the relationships between these neurons are found across areas of cortex and species \cite{harris2015neocortical}.  The maintenance and control of such distinguishing properties in the face of perturbation is equally important, as a functional role which doesn't meaningfully persist across a consistent range of perturbations (i.e. one which lacks robustness) is effectively useless. Many empirical and computational studies have focused on the nature and mechanisms of this robustness in its many flavors, including: intrinsic neuronal excitability \cite{desai1999plasticity, barth2004alteration, hong1995activity, o2014cell, marder2014neuromodulation, remme2012homeostatic} and regulation of synaptic efficacy both as it directly relates to firing rate homeostasis \cite{turrigiano2004homeostatic, turrigiano2012homeostatic, marder2014neuromodulation, ibata2008rapid} and as addressing the inherent instability of additive Hebbian spike-timing dependent plasticity (STDP) \cite{gilson2011stability, van2000stable, kempter2001intrinsic}. The difficulty of implementing multiple concurrently active plasticity mechanisms effectively in recurrent neural networks \cite{markram1997regulation, bi1998synaptic} has lead to a relative dearth of such models with a few very notable exceptions (in particular, though not exhaustively: \cite{lazar2009sorn, zheng2013network, miner2016plasticity, litwin2014formation, vogels2011inhibitory, effenberger2015self}). In particular, the pioneering work on the SORN model demonstrated that the synergy of a mere 3 plasticity mechanisms (4 counting synaptic growth/pruning) can account for a multitude of observed features in cortical microcircuits \cite{lazar2009sorn, zheng2013network, miner2016plasticity} and very much paved the way for the work detailed here.  Indeed the core of MANA's mechanisms are inspired by the SORN, due in part to the demonstration of their benefits and stability in previous work\cite{lazar2009sorn}. 

In particular, work on the SORN has demonstrated that a wide array of circuit features and behaviors can be self-organized entirely via approximations to well known plasticity rules when the distribution of TFRs is hand-tuned to a lognormal distribution \cite{miner2016plasticity}. Additionally, the dynamics of excitatory (Exc.) \textrightarrow inhibitory (Inh.), Inh. \textrightarrow Exc., and/or Inh. \textrightarrow Inh. synapses are often fully or partially ignored depending upon the self-organizing model in question \cite{lazar2009sorn, zheng2013network, miner2016plasticity, effenberger2015self, litwin2014formation}. It should be noted that iSTDP in its various flavors was excluded in these studies by design in order to focus on the investigation of other self-organizing mechanisms and was not in any way an oversight. However this clearly points toward the inclusion of iSTDP in a complete sense as a logical next step forward in the development of this class of models, and indeed without work investigating the other mechanisms this next step would not be possible. In order to self-organize from a null initial state we require rules for the dynamics of inhibitory synapses as well as some mechanism for self-organizing the TFRs of neurons, which stand in addition to the pre-established mechanisms underlying the SORN and SORN-like models. In the former case of inhibitory dynamics there exists literature on inhibitory STDP (iSTDP) from which we can draw upon for the model's inhibitory dynamics \cite{bell1997synaptic, woodin2003coincident, d2015inhibitory, fino2008cell}(for a review see: \cite{kullmann2012plasticity}). However, in the latter case, while there has been work regarding the necessary conditions for lognormal firing rates \cite{roxin2011distribution}\cite{koulakov2009correlated}, and putative rules for achieving them \cite{koulakov2009correlated}\cite{effenberger2015self} there exists no such literature on mechanisms for the evolution of the set-points of firing rate homeostasis specifically. We introduce such a mechanism: a metaplastic rule for the evolution of the set points of homeostatic plasticity and this metaplastic rule constitutes the ``M'' in MANA. 

In spite of the progress in modeling homeostatic mechanisms, very few models have focused on the second piece of the self-organization puzzle: differentiation, or how exactly the set points that homeostasis aims to achieve come about. From a purely logical standpoint one can observe that in order for homeostatic mechanisms to exist in the first place there must be a point (or set of points or manifold) in the neuron's state space which the mechanism in question makes robust to perturbations. Such is intrinsic to the notion of homeostasis. Likewise, many models (computational and conceptual) assume such set points \cite{o2014cell, turrigiano2007homeostatic, davis2006homeostatic, lazar2009sorn, zheng2013network, miner2016plasticity, effenberger2015self}, but to date very few models have studied how such set points are arrived at, the effect of their transient instability, or otherwise included them in a self-organizing model, with the notable exception of work by Yann Sweeney and colleagues \cite{sweeney2015diffusive}. In \cite{sweeney2015diffusive} NO\textsuperscript{+} concentration and diffusion was used as a means of signaling activity between nearby cells, thus providing a means of adjusting the firing rates based on the firing rates of spatially proximal cells. This mechanism, similar in concept to meta-homeostatic plasticity detailed here, though rooted in biology has yet to be validated in empirical studies. Considering the current state of the field with respect understanding how neurons come to possess characteristic firing rates, we opted to phenomenologically model this unknown mechanism in a way which most accurately reproduced specific findings in cortical tissue. In doing so we offer up meta-homeostatic plasticity as a possible mathematical and phenomenological description of the mechanism(s) which tune target firing rates and believe our work to be complementary with \cite{sweeney2015diffusive} with respect to understanding this aspect of cortical circuitry. The formulation of such a rule as presented here is, then, a possible logical next step forward for this class of model.

\section*{Materials and Methods}
\label{sec:Methods}

All simulations used Simbrain 3.0 (http://simbrain.net, \cite{tosi2016simbrain}) as a library for most basic neural network functions, with custom source code written for more esoteric features of the model.

Without initial recurrent connections the model requires some sort of external drive in order to self organize. To this end the 24 tokens used in Jeffrey Elman's 1993 paper on grammatical structure and simple recurrent networks \cite{elman1991distributed} were each converted to 100 distinct Poisson spike trains of a duration of 200 ms. These tokens were then arranged them according to the rules of the toy grammar from the same paper. The grammar includes a significant amount of temporal dependencies up to several words apart (also from \cite{elman1991distributed}). 

\begin{figure}[!h]
\includegraphics[width=0.6\linewidth]{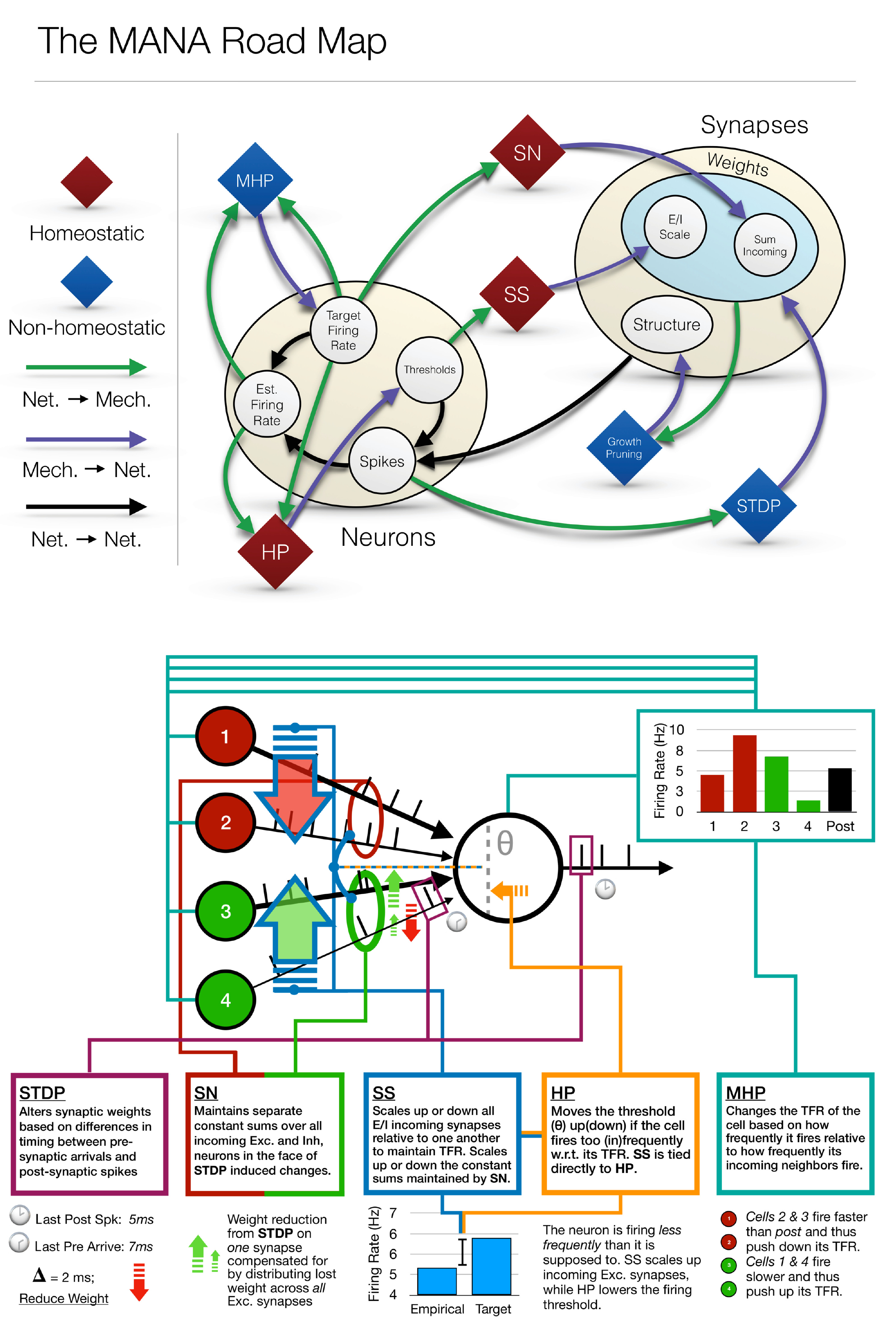}
\caption{{\bf A map of MANA's various mechanisms}
 \footnotesize A map of the interplay between network attributes and mechanisms. This map can be used as a quick reference and high-level overview of what mechanisms, MANA employs, what aspects of the network they act upon and are influenced by and how direct or indirect the actions of any one mechanism are on another. Acronyms are as follows: \textbf{MHP:} meta-homeostatic plasticity (how TFRs change), \textbf{HP:} homeostatic plasticity (how neurons alter thresholds to maintain TFR), \textbf{SN:} synaptic normalization (how neurons maintain a constant incoming total Exc./Inh. current across all afferent synapses), \textbf{STDP:} Spike-timing dependent plasticity (how synapses change strength in response to pre- and post-synaptic spikes), \textbf{SS:} synaptic scaling (how neurons shift and scale their Exc./Inh. balance to help maintain the TFR), and \textbf{Growth/Pruning:} self-explanatory (how synapses are removed from or added to the network).  Green arrows indicate where aspects, attributes, or properties of the network are used as parameters or variables for plasticity mechanisms. Purple arrows indicate action of a plasticity mechanism on a variable, and black arrows indicate direct influence between network attributes/variables. The direct influences (in the form of directly altering or directly being used as an input to a function) are all clearly visible, but perhaps more interesting is that this map can be used to trace all indirect influences. For instance spikes/spike times are the parameters for STDP, which alters synaptic weights, which in turn affect spiking, which are used to estimate the firing rates, which are used as compared to TFRs to alter thresholds, and so on. Below is a diagram of each of the mechanisms including both generic descriptions and contextual descriptions with respect to specific aspects of the cell they are responding to. 
 }
\label{Fig1}
\end{figure}

While much less complicated than a real living cortical circuit MANA is still considerably more complex than other models in a similar vein as a result of its all-inclusive goals. While this section as a whole gives a detailed account of its mechanisms, Fig. \ref{Fig1} provides a high-level overview that many readers may find convenient. 

\subsection*{Neuron and Synapse Models}

Simulations consisted of 924 recurrently connected, single-compartment, leaky integrate-and-fire neurons with firing rate adaptation.  This dynamical ``reservoir'' was driven by 100 input neurons which lacked any internal dynamics and received no back-connections from the 924 reservoir neurons, for a total of 1,024 neurons. Connections from the inputs to the recurrent reservoir were initialized to a sparsity of 25\%, had their weights drawn from $\mathcal{N}(\mu = 3;\,\sigma = 1)$, and their delays drawn from a uniform distribution [0.25, 6] ms. All input neurons in the model are excitatory and thus any weight values less than zero had their sign flipped. Input synapses behaved in exactly the same manner as reservoir synapses and were subject to all of the same plasticity mechanisms including growth and pruning. Reservoir neurons (hereafter referred to simply as ``neurons'') were modeled as leaky integrate-and-fire neurons with adaptation and were updated using the following:

\begin{gather}
C_m \frac{dV}{dt} = (V_l - V) + I_e - I_i + I_{bg} + I_{noise} - w \notag  \\
\tau_w \frac{dw}{dt} =  -w \\
V \;>\; \theta \;\;?\;\; V \leftarrow V_{reset}; \;\; w \leftarrow w + b  \notag
\label{mainNeuronEq}
\end{gather}

Where \emph{V} is the membrane potential, \emph{V\textsubscript{l}} is the leak reversal (-70 mV), \emph{w} is the adaptation current, and dot-notation is being used to denote derivatives. Spike-frequency adaptation was incremented by \emph{b} (15 nA and 10 nA for excitatory and inhibitory neurons respectively) and the membrane potential was set to the reset value (-55 mV; where \textleftarrow indicates assignment) whenever an action potential occurred. Spike-frequency adaptation and decayed with time constant $\tau_w$ of 144 ms. Neurons generated an action potential (spike) if their membrane potential exceeded their firing threshold ($\theta$), which was initialized to -50 mV (variable, governed by HP, see Subsection: \nameref{sec:HP}). All neurons had a refractory period (3(2)ms for excitatory(inhibitory) neurons) during which the membrane potential was held constant at the reset value ($V_{reset}$) and no action potentials could be generated. The membrane capacitance ($C_m$), was drawn from $\mathcal{N}(26, 1.5)$ ($\mathcal{N}(23, 2.5)$)nF for excitatory(inhibitory) neurons. $ I_{bg}$, and $I_{noise}$ are the synaptic input, background, and noise currents, impinging on the cells. 

Neurons were uniformly randomly embedded within a rectangular prism in 3-D space. Distance was not tied to any specific unit and merely existed as a value from which to derive synaptic delays. For all recurrent \textrightarrow recurrent synapses delays were proportional to distance between pre- and post-synaptic neurons in the prism, averaging 2.5 ms, with values as low as 0.5 ms and as high as 6 ms.

\begin{tabular}{ l l l l }
	\textbf{Neuron Parameters} & Exc.(Inh.)& & \\ \hline
	\emph{C\textsubscript{m}} & $\mathcal{N}(26, 1.5)$($\mathcal{N}(23, 2.5)$)nF & \emph{V\textsubscript{l}} & -70 mV \\
	 \emph{V\textsubscript{reset}} & -55 mV &  \emph{I\textsubscript{bg}} & 18.5 nA \\
	  \emph{I\textsubscript{noise}} & $\mathcal{N}(0, 0.1)$ & $\tau_w$ & 144 ms \\
	  $\theta$(initial) & -50 mV & \emph{b} & 15(10) nA \\
\end{tabular}

\subsection*{Firing-Rate Mechanisms}
	
The cornerstone of MANA is it's 2nd order plasticity mechanism (meta-homeostatic plasticity (MHP)) which changes TFRs using a local rule. However, in order to maintain or change a TFR the cell requires some sort of mechanism for determining its average depolarization rate over some time-scale. Average intracellular calcium would seem to fill this role nicely \cite{golowasch1999network}\cite{lemasson1993activity}, and although \cite{turrigiano2004homeostatic} points out that its exact role with respect to homeostatic plasticity has not been established, it has been used effectively for maintenance of depolarization dynamics in single-compartment Hodgkin-Huxley model neurons \cite{o2014cell}. Here an exponential rise and decay function was used as a proxy for a running average to allow the cell to estimate firing rate:	

\begin{gather}
	\tau_\kappa = 10^4/\sqrt{\bar{\nu}} \notag \\ 
	\frac{d\kappa}{dt} \;=\; \frac{-\kappa}{\tau_\kappa} + \delta(t - t_{spk}) \\
	\frac{d\nu_{\kappa}}{dt} \;=\; -\frac{\kappa}{\tau_\kappa} - \nu_{\kappa}, \;\;\; \hat{\nu} \leftarrow 1000\cdot\nu_{\kappa} \notag
\end{gather}
		
Notably, the rate of rise and decay is tied to the (dynamic) TFR($\bar{\nu}$) of the cell. The dependence of the time constant $\tau_\kappa$ on TFR gave parity between high and low activity neurons. In the former case the estimated firing rate (EFR) will increase more with each spike, but more quickly decay, while in the latter case, the instantaneous effect of individual spikes is diminished, but they are also less quickly forgotten. This is ideal since by definition a neuron which fires quickly must do so at least fairly regularly, while on the other hand the nature of being a low activity neuron is such that activity is spread over long periods of time. Similarly, this minimizes the impact of how firing rates are estimated on the dynamics of individual neurons. For instance, uniform application of a large $\tau_\kappa$ would bias high firing rate neurons toward bursting more than low firing rate neurons due to the longer amount of time spikes are ``remembered''. Here $\nu_\kappa$ refers to the raw firing rate estimate (in kHz), while $\hat{\nu}$ is the final firing rate estimate in Hz ultimately used in the HP and MHP terms.

%% TODO: Insert figure demonstrating how estimated firing rate tracks with avg. firing rate.

\subsubsection*{Homeostatic Plasticity}
\label{sec:HP}
Findings from \cite{barth2004alteration} indicated that neurons (regardless of fosGFP gene expression, associated with higher firing rate cells) significantly altered their membrane thresholds for spike generation. Furthermore other self-organizing models have also used alterations to firing threshold as their primary firing-rate homeostasis mechanism \cite{lazar2009sorn}. In our model, HP acted upon the neuron's firing threshold primarily, $\theta$ as well in the following manner:  	
	 \begin{gather}
	 \lambda_{hp}\frac{d\theta}{dt} = ln\left(\frac{\hat{\nu}+\epsilon}{\bar{\nu} + \epsilon}\right)
	 \end{gather}

Where $\lambda_{hp}$ is the HP constant which is initialized to $10^4$, and increased to $10^5$ by the end of the simulation exponentially with a time constant of $5 \times 10^{-6}$ ms. This was to allow TFR more direct influence on EFR early in the simulation, since the former had lognormal biasing (see Sec. \nameref{sec:MHP}) . Making changes in threshold dependent upon the logarithmic difference between the EFR and TFR term was designed reflect a proportional representation of the difference between estimated and target firing rate. That is, for a neuron with a TFR of 10 Hz the homeostasis equation alters the threshold equally for an EFR of 100 Hz or 1 Hz, moving the $\theta$ up or down respectively to maintain homeostasis. This allows neurons to fluctuate about their TFRs somewhat without the threshold changing too rapidly in response in a manner that better reflects their behavior. If the EFR of a neuron with a TFR of 50 Hz fluctuates by ~1 Hz, changes in the threshold should reflect this rather minor fluctuation relative to the neuron's TFR, by itself changing rather slowly. Too rapid a response in this context could lead to over-correction and instability. However if a neuron with a TFR of 2 Hz has an EFR of 3 Hz that is a significant departure from the neuron's average firing rate/TFR and the threshold should move to correct what amounts to a 50\% increase in firing rate accordingly. Furthermore given the likely lognormal distribution of cortical firing rates it can be reasonably assumed tht compensatory mechanisms may be operating in log- as opposed to linear firing rate space. While eventually HP will silence the cell in reaction to the overabundance activity, this gives the cell more freedom to become especially active for some (perhaps salient) specific input.  

\subsubsection*{Meta Homeostatic Plasticity}
\label{sec:MHP}

The various formulations of firing rate homeostasis imply that for each neuron there exists an individual or range of TFRs, deviations from which activate homeostatic compensatory mechanisms. To date a major component missing from extant self-organizing models has been some mechanism whereby the sets points of homeostatic plasticity are self-organized. This is due in part to the lack of an empirically observed, mathematically rigorous description of the phenomena of homeostatic set-point organization, along the same lines as--for instance--synaptic plasticity and STDP. Also problematic is the inherent possibility for extreme instability presented  when the set-point of a homeostatic mechanism is allowed to be dynamic. This is further complicated by the constraint that neurons in living neural networks can differ in average firing rate by orders of magnitude and that the overall distribution of firing rates across populations has been consistently well fit by lognormal distributions in particular \cite{hromadka2008sparse, mizuseki2013preconfigured, buzsaki2014log, nigam2016}. 

While a well-formulated empirical account of this phenomena is missing, differences between high and low activity neurons in terms of their relationships to other neurons and gene expression have been documented. The immediate early gene (IEG) c-fos is well correlated with increased activity in vivo, for instance \cite{yassin2010embedded} , and sustained elevated spiking activity has been shown to drive the expression of c-fos \cite{luckman1994induction, fields1997action, schoenenberger2009temporal}. Specifically expression of c-fos always \emph{follows} increases in spiking activity and appears to be signaled by increases in intracellular calcium following influx through voltage dependent ion channels \cite{luckman1994induction}.  In some cases this expression can occur as a result of neural firing alone \cite{fields1997action}\cite{schoenenberger2009temporal}, while in others it has been demonstrated that elevated activity is insufficient for expression of c-fos, which can only occur if the elevated neuronal firing is a result of increased synaptic activity \cite{luckman1994induction}. While sensory deprivation does not diminish the presence of c-fos expressing neurons, it does diminish the differences in wiring between c-fos positive and negative neurons \cite{benedetti2012differential}.

Meta-homeostatic plasticity (MHP) introduced here, represents a phenomenological account of how neurons self-organize their homeostatic set-points, i.e. their TFRs, which is both stable and produces lognormal distributions of target and empirical firing rates. The rule is loosely based upon the known relationships between elevated neuronal and synaptic activity and the expression of c-fos, where it is hypothesized that c-fos acts in some way as a marker, indicator, or direct instantiation of a TFR variable or the process that governs it. However it should be reiterated that since this process is not fully understood the mechanism here is phenomenological in nature, merely providing a possible means by which TFRs can evolve in a stable way that results in a lognormal distribution over the population. Models including homeostatic plasticity need set-points and MHP provides a means of allowing those set-points to be self-organized in a manner producing realistic results. 

\begin{figure}[!h]
	\centering
	\includegraphics[width=\linewidth]{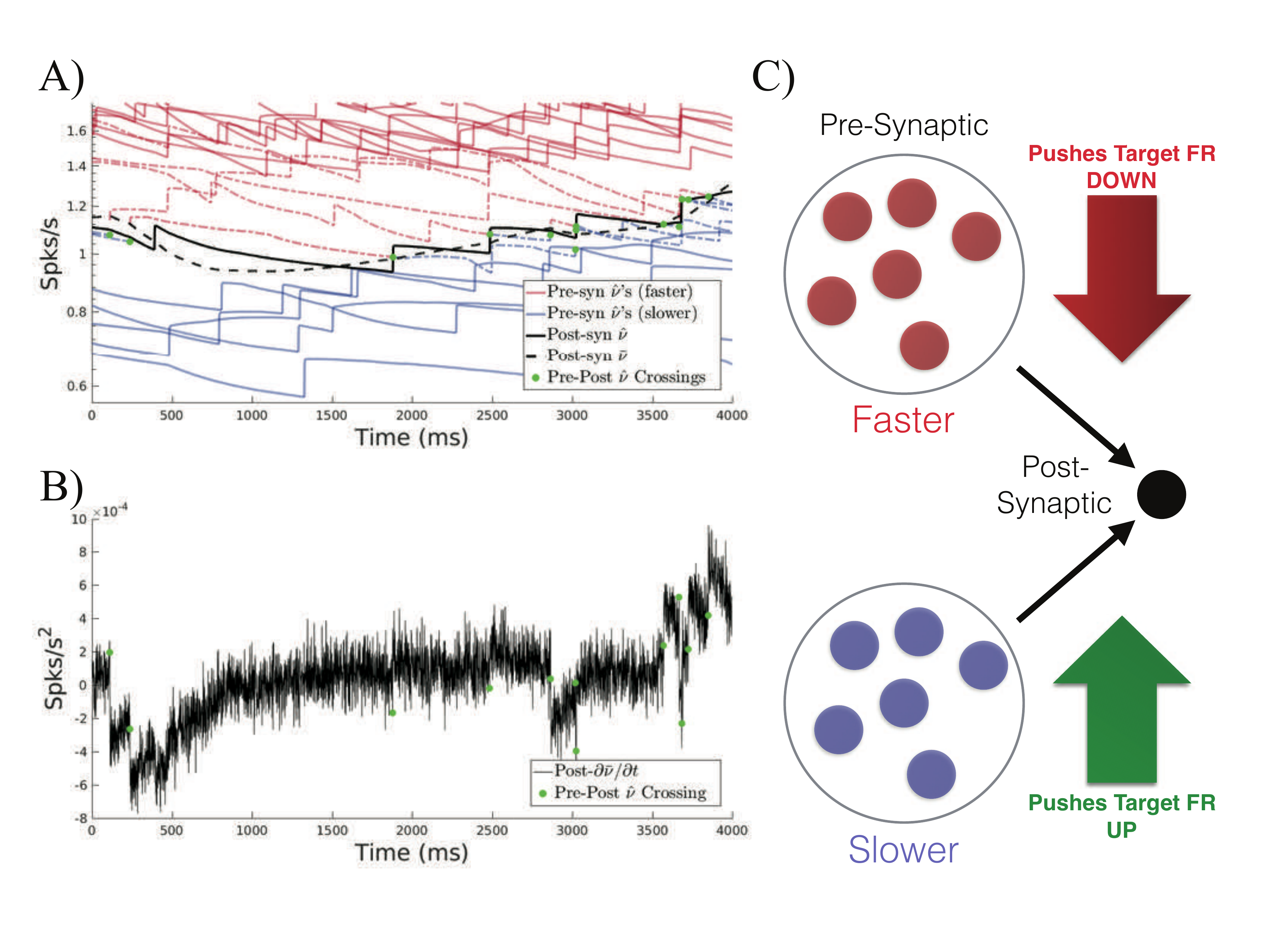}
	\caption{{\bf Meta-homeostatic plasticity Diagram. }
		 A) TFR evolves over time based on the relationship between the EFR of a given neuron and the EFR of its incoming neighbors. Here the EFRs of a neuron's nearby (in firing rate space) in-neighbors are plotted over time such that those with greater EFRs than the post-synaptic neuron are in red for all times where they are greater and blue otherwise. Green dots indicate points where the EFR of the post-synaptic neuron (black) crosses from above or below the EFR of one of its in-neighbors. The TFR of the post-synaptic neuron is represented by the black dashed line. More significant changes in direction can be noticed near green crossing points. B) A trace of the derivative of the post-synaptic neuron's TFR over the same time period as (A), and with green dots at the same points in time as in (A) indicating EFR crossings. Notice sharp changes after crossings. C) A simplified diagram of MHP: Pre-synaptic (in-neighbors) with higher EFRs than the post-synaptic neuron exert a downward force on post-synaptic TFR, while in-neighbors with lower EFRs push their post-synaptic neighbors' TFRs up. This repulsive force drops off with distance in EFR-space and is scaled so as to ultimately induce a lognormal distribution across the population. }
	\label{Fig2}
\end{figure}

MHP uses the following formulation, which assumes that TFRs evolve based on local firing rates and the firing rates of incoming neighbors. The relationship between a neuron's firing rate and the firing rates of its in-neighbors is such that in-neighbors with lower firing rates exert a positive force while in-neighbors with higher firing rates exert a negative force. This repulsive force decays based on the difference between pre- and post-synaptic firing rates such that the greatest positive or negative force is exerted by in-neighbors whose firing rates are very close to that of the post-synaptic cell. Alternatively, this can be thought of from the perspective of the in-neighbors, and by this token every neuron in the network can be viewed as pushing the TFRs of their out-neighbors with similar levels of activity away based on their own firing rate at any given moment. In this way changes in the sign of the derivative of TFR (or large changes in general) are precipitated by the firing rate of a  post-synaptic neuron crossing above or below the firing rate of one of its in-neighbors (thus changing the sign of the contribution of that in-neighbor to MHP). This can be seen in Fig. \ref{Fig2} which displays the EFRs of the in-neighbors of a given neuron superimposed over the EFR of the post-synaptic neuron as well as how this effects the neuron's TFR and its derivative. Thus in this formulation, a neuron's TFR evolves as a function of how its empirical firing rate relates to the firing rates of its in-neighbors, with the latter constraining its evolution within the context of its neighbors in the network. This prevents over-synchronization or ``clumping'' of many neurons around the same firing rate, which is difficult to prevent if local empirical firing rate (and no explicit information about the firing rates of other neurons in the network) fully governs the evolution of TFR. 

In more formal terms: For each neuron $j \in \{1, ... , N\}$ (where $N$ is the number of neurons in the network) there exists a set of neurons $I_j(t)$ consisting of the $M \in [0, N]$ neurons which send synaptic projections to $j$ at time $t$. This gives the set  $I_j(t) \coloneqq \{\,i = 1, ...\, , M \;|\; w_{ij}(t) \neq 0\,\}$. Along the same lines as work done in \cite{gilson2011stability} on STDP, we use the Fokker-Planck formalism to study (and reason on how to influence) the probability density $\mathcal{P}(J)$ of a population of TFRs \emph{J}, which in our case are modified by many continuous interactions with presynaptic neurons (as opposed to discrete plastic updates as with STDP). That is, we regard the TFR of an individual neuron $j$, $\bar{\nu}_j \in J^p$ at time $t$ to be the sum of modifications caused by interactions with presynaptic neurons: $\bar{\nu}_j(t_f) = \int_{0}^{t_f} f(\bar{\nu}_j(t), I_j(t); \theta) dt$. Where $\theta$ represents some non-specific set of parameters. As in \cite{gilson2011stability} we observe that there exists a family of functions for the drift ($A(J)$) and diffusion ($B(J)$) of \emph{J} for which the distribution produced by the stationary solution of the Fokker-Planck equations is approximately lognormal. Namely:
\begin{gather}
	\exists A,B \;\;F(A),\,F(B) \text{  where: } \notag \\ \notag\\  \frac{\mathcal{N}}{B(J)}  exp \left[\int_{0}^{J }  \frac{2A(J^\prime)} {B(J^\prime)} d(J^\prime)\right]\propto \frac{1}{\sqrt{2\pi}sJ} exp\left[-\frac{[ln(J)-m]^2}{2s^2}\right] \notag \\ 
\end{gather}

From this two biasing functions $f_+/f_-$ for STDP were derived in \cite{gilson2011stability}. Considering the similarities between a synaptic weight defined by the sum of multiple updates in STDP and TFR defined by the sum of multiple updates here the  $f_+/f_-$ terms derived in \cite{gilson2011stability} are used to act on TFR instead of synaptic weight. Equations (5-6) were chosen so as to satisfy the assumptions of the derivation in \cite{gilson2011stability}.

 We further define two mutually exclusive subsets $L_j(t) \subseteq I_j(t)$ and $G_j(t) \subseteq I_j(t)$ where $L_j(t) \coloneqq \{\,i \in I_j(t) \;|\; \hat{\nu}_i(t) < \hat{\nu}_j(t)\,\}$ and $G_j(t) \coloneqq \{\,i \in I_j(t) \;|\; \hat{\nu}_i(t) > \hat{\nu}_j(t)\,\}$. That is, $L_j(t)$ ($G_j(t)$) is the subset of neurons projecting onto neuron $j$ which have lower (higher) EFRs than $j$ at time $t$. The specification of time being necessary in our definitions since activity levels (and ergo set membership), TFR, and even membership in $I_j(t)$ (as a result of synaptic pruning) are all dynamic. In order to explicitly prevent ``clumping'' of neurons around the same preferred activity level the EFRs of neurons in $I_j(t)$ exert a \emph{repulsive effect} with respect to $\hat{\nu}_j$ which influences $\bar{\nu}_j$. Specifically neurons in $L_j$ (less active than $j$) produce a potentiating effect on  $\bar{\nu}_j$ while neurons in $G_j$ (those more active than $j$) depress \emph{j}'s TFR according to the following rule:
 
 \begin{gather}
	 \frac{\partial \bar{\nu}_j}{\partial t} \;=\; \eta \frac{1}{|I_j|} \left[ f_+(\bar{\nu}_j) \left(\sum_{i \in L_j(t)} e^{-|u|}\right) - f_-(\bar{\nu}_j) \left(\sum_{i \in G_j(t)} e^{-|u|}\right) + \zeta \right]  \\ \notag \\
	 u = \frac{\hat{\nu}_j - \hat{\nu}_i}{\bar{\nu}_j} \\ \notag
 \end{gather}

Where $\eta$ is the learning rate which was initialized to 0.05 but exponentially decayed to 10\textsuperscript{-6} with a time constant of 500s. Note that $\eta$ was set to 0 when a neuron reached its maximum incoming inhibitory and excitatory currents (see: \nameref{SN}). The contributions of incoming neighbor neurons to the change in TFR are also normalized by the in-degree of the neuron i.e. the instantaneous set cardinality of \emph{I\textsubscript{j}}, denoted by $|.|$. $\zeta$ here is a noise term drawn from $\mathcal{N}$(0, .7). As \cite{gilson2011stability} has derived $f_+/f_-$ equations which demonstrably represent an approximately lognormal solution to the stationary Fokker-Planck equations, and shown them to be successful in the context of (Log-)STDP, the same terms are reused here:
		 
\begin{gather} 
	 f_+(\bar{\nu}) \;=\; e^{-\bar{\nu}/\beta \nu_0} \\
	 f_-(\bar{\nu}) \;=\;
	 \begin{cases}
	 \bar{\nu}/\nu_0, &  \bar{\nu} \leq \nu_0 \\
	1 + \frac{log\left(1 + \left(\alpha \cdot (\bar{\nu}/\nu_0 -   1)\right)\right)}{\alpha}, & \bar{\nu}> \nu_0 
	 \end{cases}\\ \notag
\end{gather}
		
Where $\alpha$ determines the degree of log-like saturation i.e. if firing rate depression for a neuron with a $\bar{\nu}$ above $\nu_0$ (the "low" firing rate constant, set to 2 Hz) has a large $\alpha$ the depression relative to $\bar{\nu}$ will be more log-like and likewise will be closer to linear for small $\alpha$. $\beta$ controls the rate at which $\bar{\nu}$ facilitation decreases with increasing $\bar{\nu}$, such that small $\beta$ entails a rapid fall-off in firing rate potentiation, and large $\beta$ entails a slower fall-off. In all simulations, $\alpha$ and $\beta$ were set to 2.5 and 10, respectively. An example of how local EFR, the EFR of incoming neighbors, and the TFR interact can be seen in Fig. \ref{Fig2}, where changes in the derivative of TFR can be seen changing in response to the post-synaptic neuron's EFR moving above or below the EFRs of one of its pre-synaptic neighbors (resulting in a change in the sign of the contribution of that pre-synaptic neuron's EFR).
	\newline \newline
\begin{tabular}{ l l l l }
	\multicolumn{4}{l}{\textbf{MHP \& HP Parameters}}  \\ \hline
	$\lambda_0$ & $10^4$ & $\lambda_f$ & $10^{5}$ \\
	$\eta_0$ & 0.05 & $\eta_f$ & $10^{-6}$ \\
	$\alpha$ & 2.5&  $\beta$ & 10 \\
	$\nu_0$ & 2 Hz & $\zeta$ & $\mathcal{N}(0, .7)$ \\
\end{tabular}
	
\subsection*{Synaptic Plasticity}
	
Long-term synaptic plasticity was driven by spike-timing dependent plasticity (STDP) and synaptic scaling (normalization, SN). These alterations to synaptic strength were the primary driving factor behind the resulting connectivity structure, the other being the pruning protocol (see Subsection: \nameref{subsec:prune} ). While on the surface it may appear that the pruning rules bear the bulk of the responsibility for the resulting connectivity, only weak synapses are eligible for pruning, and it is STDP and SN which determine the relative strength of a synapse and thus its eligibility for said pruning. 
	
All weight changes (be they through STDP or SN) were put through the following dampening function which prevented any synaptic weight $w$ from exceeding some maximum weight $W_{max}$ by reducing any potentiation or depression commensurate with $w$'s proximity to $W_{max}$.

\begin{gather} \label{eqn:8}
	\small
	\ f_{damp}\left(w; \Delta w'\right) \,=\, \Delta w'\left(\frac{1-e^{\left(\frac{w}{W_{max}}\right)^2 - 1}}{1-e^{-1}}\right)\\
	\Delta w \;=\; f_{damp}(w; \Delta w') \notag
\end{gather}	
		
Here $\Delta w'$ refers generically to any change to a synapse's weight (discrete or continuous) and thus all future references to weight changes should be considered as having been passed through this function. In our model $W_{max}$ was set to +/- 200 nA. This is as opposed to $\Delta w$, which is the actual applied change. This sort of dampening has been observed in existing synapses, for instance \cite{bi1998synaptic} found that STDP induced LTP had little effect on already strong glutaminergic synapses in dissociated hippocampal cultures and that fluctuations in spine volume between pyramidal cells in cortex were reduced in either direction for more strongly coupled neurons. Further in living tissue synapses cannot become infinitely strong and there are limits to the amount of neurotransmitter they can release and sensitivity of the post-synaptic cell to that neurotransmitter.		
		
\subsubsection*{Short-Term Plasticity}

Post-synaptic currents (PSCs) were modeled as an instantaneous jump and decay, with dynamic jumps representing short-term plasticity as modeled by the Use, Depression, Facilitation (UDF) model \cite{markram1998differential}.

\begin{gather}
u \;=\; U + u(1-U)e^{\Delta_k/F}  \\
R \;=\; 1 + (R - uR - 1)e^{\Delta_k/D} \notag \\
A_k \;=\; R \cdot w_k \cdot u\notag \\
\frac{dq_{psr}}{dt} = -q_{psr}/\tau_{psr} \,+\, A_k\cdot\delta(t - t_{arr}) \notag 
\end{gather}

The UDF model is designed to account for variations in post-synaptic response caused by depletion of neurotransmitter (depression) and influx of calcium between spikes (facilitation) \cite{tsodyks2013short}. The synaptic parameters U (use), D (depression time constant), and F (facilitation time constant) were fixed for each synapse, being drawn from different normal distributions depending on whether the pre- and post-synaptic neurons were excitatory or inhibitory. The mean values for U, D, and F (with D and F expressed in seconds) were: .5, 1.1, .05 (EE), .05, .125. 1.2 (EI), .25, .7, .02 (IE), and .32, .144, .06 (II), with standard deviations set to half the mean, and negative values re-drawn from the distribution until positive. This is consistent with \cite{markram1998differential}, and uses the same parameters found in much of the liquid state literature (when UDF is included, e.g. \cite{maass2002real}). Here $\Delta_k$ is the most recent inter-spike interval (ISI) for neuron k, where the ISI is calculated as the difference between the last spike arrival and the arrival of the current spike. The value $w_k$ represents the strength or weight of outgoing synapse $k$. In the final equation $q_{psr}$ is taken to be the total post-synaptic response impinging on synapse $k$'s post-synaptic neuron, and where $\tau_{psr}$ is a decay time constant set to 3(6)ms for excitatory(inhibitory) pre-synaptic neurons. Lastly $\delta(t - t_{arr}) $ is the Dirac-delta function of the current time subtracted from the arrival time of the pre-synaptic spike at the post-synaptic terminal. This is not the same as the spike time of the pre-synaptic neuron due to synaptic delay. 
		
\subsubsection*{Synaptic Normalization}
\label{SN}
	
Neurons in the model took steps to ensure that the sum of incoming synaptic currents was kept at a constant value, unique to each neuron. In self-organizing models with a constant homeostatic mechanism (one which favors the same firing rate across all neurons) as implemented in \cite{lazar2009sorn, zheng2013network, miner2016plasticity} a constant synaptic normalization sum is sensible. Along those lines, one could imagine producing any desired distribution of firing rates in a network solely through manipulation of thresholds, even if total synaptic input were held constant across the constituent neurons. But, while such a scenario is possible in principle it places too much responsibility upon manipulation of the threshold, which must fight \emph{against} this undue homogeneity of synaptic inputs imposed by our hypothetical modeler. Furthermore, from the standpoint of realism it is known that higher firing rate neurons tend to receive more total synaptic connections than their slower counterparts \cite{yassin2010embedded}\cite{benedetti2012differential}, and studies using transfer entropy have demonstrated a high degree of inequality among neurons in terms of information flow with some 20\% of neurons accounting for 70\% of information flow in vitro \cite{nigam2016}. 

Attempts to include this sort of heterogeneity have appeared in other models, perhaps most notably \cite{effenberger2015self} where the incoming sum to be normalized was tied to excitatory synaptic in-degree. This produced interesting dynamics including the emergence of excitatory driver cells. However, such a configuration makes the total input to each cell predetermined by the modeler and does not allow cells to develop in accordance with the history of the network in which they are embedded.Fortunately there exists an explicit variable here which is itself dynamic, plastic, and otherwise self-organized which can be used for the purpose of determining total input to each neuron: target firing rate ($\bar{\nu}$). 
		
	\begin{gather}
		 \overset{\infty}{I}_{e/i}(t) = \frac{\bar{\sigma}_{e/i}\omega_a}{1+\text{exp}(-\omega_b\bar{\nu}(t))} - \omega_c + inp_0 \\
		 \overset{\infty}{I}_{e/i}(t) > k_{in}W_{max}\;?  \;\;\;
		 \overset{\infty}{I}_{e/i}\leftarrow k_{in}W_{max} 
	\end{gather}
	\label{Eqn:16}
		Normalization proceeding as follows: for each $w_{ij}$ where $ i = \{1, 2, ... N\}$ and j is the index of the target neuron:
		\begin{gather}
			\Delta w_{ij} = \overset{\infty}{I}_{e/i}\frac{|w_{ij}|}{\sum_{i \in S_{e/i}} w_{ij}} - w_{ij}
		\end{gather}

Here $\overset{\infty}{I(t)}_{e/i}$ is the maximum total current (saturation value) of each type allowed to impinge on each neuron at time $t$. Both currents have a linear dependence upon the cell's TFR up to a certain point. The logistic sigmoid is used here to represent the saturation of total current impinging on a cell causing an initial roughly linear dependence upon the TFR which eventually nonlinearly approaches some predetermined maximum. The shape of the sigmoid determines how high $\bar{\nu}$ can be before increases in $\bar{\nu}$ begin providing diminishing returns with respect to the total allowed current of that time. It also determines the value at which total current saturates. The three shape parameters $\omega_a$, $\omega_b$, and $\omega_c$ were set to 300, 0.1 and 100 respectively such that (not accounting for inp\textsubscript{0}) the minimum Exc./Inh. saturation for $\bar{\nu} = 0$ was 500 nA, while the maximum total Exc./Inh. current was 2,000 nA. Each neuron's saturation 
had an additional term added: inp\textsubscript{0} which was the sum of inputs from the input layer. This gave each neuron an equal capacity  in terms of synaptic input to each neuron which could be populated by the recurrent connections which were expected to grow. On average inp\textsubscript{0} was ~750 nA. One may consider this a substantial bias, after all some neurons would be initialized with higher saturation values for all values of $\bar{\nu}$ than others. However, in practice heterogenous initial inp\textsubscript{0} (both in terms of the higher allowed saturation and all around more input from the input layer) did not bias the network in this way and inp\textsubscript{0} was a poor overall predictor of final $\bar{\nu}$.
		
\subsubsection*{STDP}
				
STDP operated on all types of synapses: EE, EI, IE, and II. Where EE refers to a synapse from an excitatory neuron to another excitatory neuron, EI refers to a synapse from an excitatory neuron to an inhibitory neuron, and so on. Different STDP windows were used for each case, since STDP observed at synapses involving inhibitory neurons (as either the pre- or post-synaptic cell) can take on a multitude of different forms \cite{rubenstein2013neural}\cite{vogels2013inhibitory}. STDP used a small learning rate and updated weights continuously rather than in discrete jumps. This diminished the effect of repeated instances of spike time pairings, but overall was motivated by the fact that since the MANA reservoir starts with (effectively) no connectivity continuous growth seems more logical.

For EE STDP a standard Hebbian window was chosen \cite{bi1998synaptic}. This window was also chosen for EI STDP as found in \cite{fino2008cell} at fast-spiking striatal interneurons. Although an investigation of effects of all the different known EI STDP windows on synaptic structure and neural activity in the context of a self organizing network would be compelling it is out of the scope of this paper. In this work, EE and EI STDP rules took on the familiar additive form as follows:
	
For $\Delta_t = t_{pre} - t_{post}$:
\begin{gather}
	\frac{dw}{dt} \;=\; \eta_{stdp}W_{+/-} e^{-|\Delta_t|/\tau_{+/-}} \\ \notag
\end{gather}
	
However, when passed through the dampening function, a common practice to keep synaptic weights within \emph{w\textsubscript{max}}, the rule technically becomes multiplicative, though unlike Log-STDP not in itself capable of inducing lognormal weight distributions \cite{gilson2011stability}. For Hebbian EE and EI synapses, LTP (LTD) occurs when $\Delta_t < 0$ ($\Delta_t > 0$), and thus $W_+/\tau_+$ ($W_-/\tau_-$) are used. In the above $w$ refers to synaptic strength while $\eta_{stdp}$ is the learning rate or time-constant of weight changes caused by STDP.
	
For all synapses emanating from inhibitory cells (Ie/iI) a symmetric Mexican-hat function (the negative 2nd derivative of the normal probability density function) was used for the STDP window, which has been found at inhibitory afferents to CA1 pyramidal neurons \cite{woodin2003coincident} and is consistent with findings from auditory cortex \cite{d2015inhibitory}. While in both cases this was only observed at IE connections, wanting to self-organize all our synaptic connectivity entailed using some STDP window for II synapses. Due to the dearth of literature as to II-STDP, the same window used for IE synapses was used as a stand in. The scaled Mexican-hat function is as follows: 
	
\begin{gather}
	\frac{dw}{dt} \;=\; \eta_{stdp} W_{+/-}\frac{2a}{\sqrt{3\sigma}\pi^\frac{1}{4}}\left[1-\left(\frac{\Delta_t}{\sigma}\right)^2\right]e^{\frac{-t^2}{2\sigma^2}} \\ \notag
\end{gather}
	
Where $a$ is a scaling factor set to 25 for both IE and II synapses and sigma determines the overall width of the window, 22 for IE connections and 18 for II connections. The choice to use the IE-STDP rule found in \cite{woodin2003coincident} and \cite{d2015inhibitory} appeared to work well, though it is apparent that the topic could use further empirical and computational investigation particularly with respect to II-STDP. 

\subsubsection*{Synaptic Scaling}

The choice to include dynamic inhibitory synapses in MANA precludes ignoring inhibitory synapses with respect to synaptic normalization. In addition to the question of how the target sums of synaptic normalization ought to interface with a network of neurons with heterogeneous TFRs, we must also consider how synaptic normalization treats inhibitory afferents. Because of this, the ratio of incoming excitation to incoming inhibition becomes another degree of freedom which requires regulation. Synaptic normalization as a mechanism has its roots in the notion of homeostatic synaptic scaling \cite{lazar2009sorn} and therefore a sort of regulation of the ratio of total incoming excitatory/inhibitory drive which behaves in a homeostatic manner follows.  It is known that neurons maintain a balance between the total inhibitory and excitatory conductances impinging upon them \cite{haider2006neocortical} , with both values scaling roughly linearly between the start and finish of UP-states. Notably neurons from \cite{haider2006neocortical} tended to maintain roughly the same slope over the course of UP-states when their total  excitatory and inhibitory conductances were plotted against each other over time. This implies that through some mechanism(s) neurons come to a roughly stable ratio of incoming excitation to inhibition. Results from \cite{benedetti2012differential} and \cite{yassin2010embedded} would seem to also back up this assertion as they noted that higher firing rate pyramidal neurons tended to receive overall less inhibition than their less excitable counterparts. Lastly it has been shown directly that brain-derived neurotrophin factor (BDNF), the production and possibly release of which is regulated by activity, decreases the amplitude of excitation between pyramidal neurons while increasing the amplitude of excitation from pyramidal neurons to interneurons \cite{rutherford1998bdnf}\cite{turrigiano1999homeostatic}\cite{turrigiano2004homeostatic}. Furthermore decreases in BDNF weaken excitatory connections onto inhibitory neurons while multiplicatively strengthening synaptic connections between pyramidal neurons \cite{rutherford1997brain}. Similarly activity blockades (resulting in the reduction of BDNF) can gobally decrease the percentage of GABA-positive neurons in vitro \cite{rutherford1997brain}. In order to both model these phenomena and provide homeostatic control over inhibition, a simple rule whereby total incoming inhibitory and excitatory currents posses independent multiplicative factors which both track with homeostatic changes in firing threshold (though in opposing directions) was used:

\begin{gather}
	\bar{\sigma}_e = e^{(\langle \theta \rangle_e - \theta)/\rho} \\
	\bar{\sigma}_i = e^{(\theta - \langle \theta \rangle_i)/\rho} \notag\\
	\langle \theta \rangle_{e/i}(t) = \frac{\Delta t \theta(t)_{e/i}}{\lambda} \,+\, \left(1 - \frac{\Delta t}{\lambda}\right)\theta(t-\Delta t) \\
	\langle \theta \rangle_{e/i} \leftarrow \langle \theta \rangle_{n} \text{;  iff } n < t_{e/i} \notag \\ 
\end{gather}

Here $\bar{\sigma}$ refers to the scaling factor which is applied to either total incoming inhibition or total incoming excitation as detailed in Sec. \nameref{SN}. $\langle \theta \rangle_{e/i}$ is an exponential running average of the neuron's threshold using the same time constant ($\lambda$) as homeostatic plasticity. Initially $\langle \theta \rangle_{e}$ and  $\langle \theta \rangle_{i}$ are exactly the same, what differentiates them is the ``trigger times'' $t_e$ and $t_i$ after which $\langle \theta \rangle_{e}$ and $\langle \theta \rangle_{i}$ respectively stop updating their values thus freezing in place. The mechanism that initiates this freeze is detailed in the next section, but in short it is the time that excitatory/inhibitory synaptic normalization starts. Before such a time synapses are growing. Synaptic normalization cannot be enforced until some condition is met, in this case the condition is whether or not total incoming excitatory/inhibitory current exceeds what those values \emph{should} be based on the synaptic normalization equations. If it were to be enforced before this point there would be little purpose in allowing the network to grow/prune its synaptic connections since the weights would immediately be scaled to sum to a specific value. Notably $t_e$ and $t_i$ can be different values because this condition is met independently for excitatory and inhibitory inputs. In any case the scaling terms are in essence the exponential difference between the current threshold and initially an exponential running average with a constant $\rho$ set to 5 in all simulations. While technically unbounded the span of all thresholds across all neurons in all simulations never exceeded 10 mV and variations within the same neuron were quite small (typically $<\pm$0.1 mV) once settled. 
	
\subsubsection*{Growth and Pruning}

\label{subsec:prune}
Synapses were initialized between every neuron in the network (all to all connectivity) and set to $10^{-4}$ nA (for context noise current impinging on the membrane potentials was drawn from: $\mathcal{N}(0, .1)$), meaning that nearly immediately a large portion of neurons took on a weight value of 0, effectively no longer existing in the network. These synapses were eventually deleted in earnest upon the first pruning cycle. Thus all synaptic connections which survive after the first cycle can be thought of as having grown from nothing, which is to say that although programatically the network is initialized to a state of full connectivity, \emph{effectively} it is initialized with no synaptic connections. All pruning cycles after the first can be conceptualized as deleting connections from the initial synaptic arbor. Each cycle was carried out at a specific interval, in this case every 5 seconds of simulated time. 
	
The pruning rule removes only the weakest synapses and does so preferentially from neurons of high-degree. Even if synapses are eligible for deletion, the probability of removal becomes smaller the lower the degree of a neuron, so as to reduce the likelihood of producing neurons which receive no connections from the excitatory and/or inhibitory neurons in the network or have no outgoing synaptic connections. This prevents neurons from being completely disconnected from the network.
	
Specifically for a synapse $s$ with an absolute efficacy of $w$ emanating from a source/pre-synaptic neuron with a set of outgoing synaptic connections $O$, and which projects to a post-synaptic neuron with sets of incoming excitatory and inhibitory synapses $S_{e/i}$. The probability of s being removed from the network is as follows: 

\begin{gather}
	p_{remove}(s_{ij}) = 
	\begin{cases}
		0 & w > \gamma\cdot w_{max} \\
		1 & w < w_{min} \\
		\frac{|S_{e/i}|}{N_{e/i}} \cdot \left(\frac{|O|}{N}\right)^2 & \text{otherwise}  \\
	\end{cases}
\end{gather}
	
Where $|.|$ refers to set cardinality, $N_{e/i}$ is the number of excitatory/inhibitory neurons in the network, and $N$ is the total number of neurons in the network. $w_{max}$ is the absolute efficacy of the strongest excitatory/inhibitory synapse in the network depending upon whether $i$ is an excitatory or inhibitory neuron, and $w_{min}$ is an arbitrary value set to $10^{-3}$ nA, which is simply meant to guarantee the removal of impossibly weak connections. Note that this value is in fact greater than the value to which synapses are initialized at the beginning of the simulation, thus growth in the first 5 seconds is a \emph{requirement} to remain in the network. $\gamma$ represents the proportion relative to the largest extant synapse such that synapses below $\gamma w_{max}$ are eligible for deletion in a probabilistic fashion. 

Synaptic growth occurred using a probabilistic quota system whereby the probability of disconnected pairs receiving a new connection between them was based upon their distance from one another in 3D space. If no synapse was created between a given pair a new unconnected pair would be selected. This occurred until the quota was filled. Two quotas existed: one for excitatory synapses and one for inhibitory synapses. The quota could not exceed more than 0.1\% of the total possible synaptic population (i.e. if 1 million synapses were possible then no more than 1000 could ever be added). 

\begin{gather}
	Q =  \begin{cases}
		Q_{max} & R \geq Q_{max} \\
		R & \text{otherwise}
	\end{cases}\\
	Q_{e/i} = \floor*{\frac{Q\cdot R_{e/i}}{R}}\\
	p_{connect}(a, b) = C \cdot e^{-\left(\frac{D(a, b)}{\lambda}\right)^2}
\end{gather}

Here $Q_{e/i}$ is our quota for adding excitatory and inhibitory synapses respectively, while $R_{e/i}$ is the total number of those synapses which were removed during the last pruning. The probability of forming a connection followed a distance based rule originally used in \cite{maass2002real}. D(a,b) is the euclidean distance in 3-space between unconnected neurons a and b, and here $\lambda$ is a regularizing parameter set such that the maximum possible distance resulted in a probability of connecting of at least 1\% before the multiplication by the constant C which was set to 0.4 in all simulations. This gave a minimum probability to connect of 0.4\%. Growth phases immediately followed pruning phases and where thus carried out at the same interval. New weights were initialized in the same manner as at the beginning of the simulation, notably with a very low efficacy. This means that newly grown synaptic connections had a negligible effect on network dynamics. Instead they served as a random detector of temporally correlated activity between the pre- and post-synaptic neuron. If no such temporal correlation existed or was too weak the synapse would fail to substantially grow and eventually be pruned, having a negligible effect on the post-synaptic neuron during its entire lifetime. Alternatively if some temporal correlation did exist and was sufficiently strong, the synapse would grow, establishing a new pathway through the network. In effect this would replace a perhaps purely correlational relationship with a potentially direct causal one. 

\begin{tabular}{ l l l l }
	\multicolumn{2}{l}{\textbf{Synapse Parameters} }& \multicolumn{2}{l} {(EE / EI / IE / II)     (Exc.(Inh.))}  \\ \hline
	U &  .5 / .05 / .25 / .32 & D & 1.1 / .125 / .7 / .144 s \\
	F & .05 / 1.2 / .02 / .06 s & $\tau_{psr} $ & 3(6) ms  \\
	\emph{W\textsubscript{max}} & 200 nA & $\gamma $ & .06 \\
	\emph{W\textsubscript{+}} & 5.1 / 5.1 / 1.8 / 1.6 & \emph{W\textsubscript{-}}  & .9 / .9 / 1.8/ 2.2 \\
	$\tau_{+} $ & 25 / 25 / - / - ms & $\tau_{-}$ & 100 / 100 / - / - ms \\ 
	a & - / - / 25 / 25 & $\sigma$ & - / - / 22 / 12 \\
	$\eta_{stdp}$ & $10^{-5}$ & $\omega_a$ & 300 \\
	$\omega_b$ & 0.1 & $\omega_c$ & 100 nA \\
\end{tabular}

% Results and Discussion can be combined.
\section*{Results}

Primary to the network's self-organization regime is the specialization of neurons, specifically their convergence upon a unique TFR and the subsequent differences in degree and neighbor preferences accompanying that value. Interestingly self-organizing TFRs seems to also lead to a differentiation of a multitude of properties across different neurons in the network. While it is true that TFR appears as a variable in other places (notably as a term in calculating maximum allowed synaptic input for Synaptic Normalization), this alone does not lend itself as an obvious answer for why certain cells developed certain differences. Our analyses can be thought of as primarily concerned with ascertaining to what degree MANA can capture features of living neural circuits with emphasis on the heterogeneity between neurons which self-organized as a result of the metaplastic mechanisms. 

\subsection*{Firing rate statistics}

\begin{figure}[!h]
	\includegraphics[width=\linewidth]{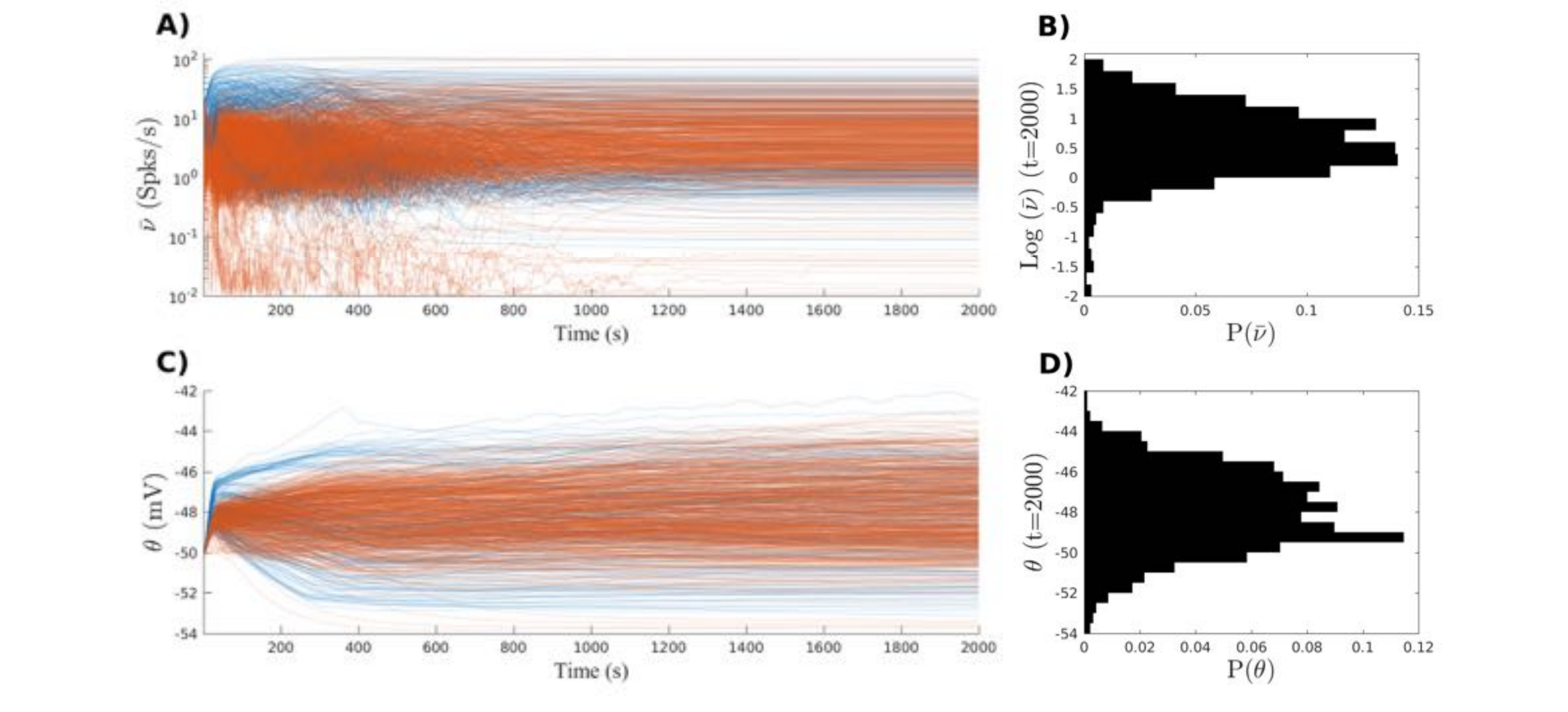}
	\caption{{\bf The evolution of TFRs and thresholds}
		\footnotesize  Here we can see how the neurons' TFRs (A) and thresholds (C) developed over the course of one of the simulations colored by the polarity of the neuron (excitatory: orange; inhibitory blue). The distributions of (A) and (C) at t = 2,000s can be seen in (B) and (D) respectively. The intrinsic plasticity mechanism begins the simulation with a very high learning rate, which decays exponentially with time loosely analogous to temperature in a simulated annealing or heat-bath algorithm. Simultaneously the homeostatic plasticity mechanism altered the firing threshold, acting as an attractive force attempting to pull the TFR into whatever value it happened to take on at the time. }
	\label{Fig3}
\end{figure}

\begin{figure}[!h]
	\includegraphics[width=\linewidth]{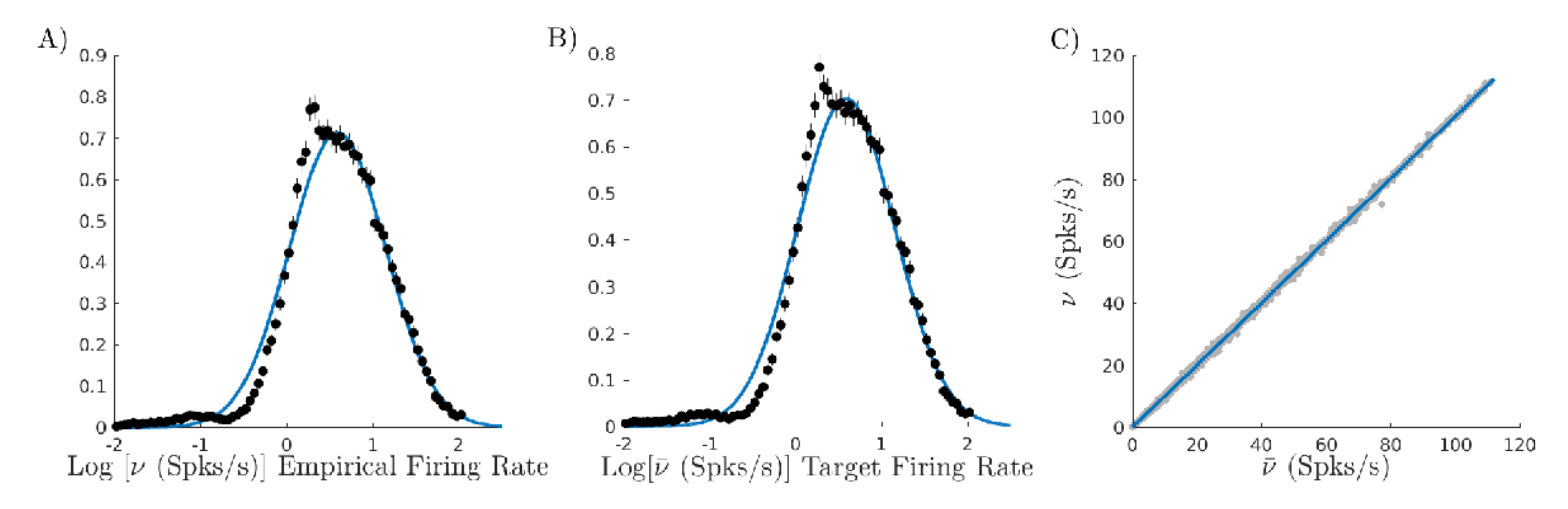}
	\caption{{\bf Roughly lognormal distribution of TFRs across 40 networks.}
	A) The distribution of empirically observed firing rates calculated by counting the number of spikes from each neuron during the last 700s of the simulation for all neurons across all 40 networks accompanied by a lognormal fit. B) A similar fit for the TFRs ($\bar{\nu}$), demonstrating that both target and empirical firing rates converge upon a roughly lognormal distribution. C) In order to ensure that MHP and HP were working properly not only do both target and empirical firing rates have to take on a roughly lognormal distribution, but for each individual neuron the TFR should mirror the average empirical firing rate. Here we plot target and empirical firing rates against each other for all neurons across all simulations and fit the results with a linear function ($R^2$ = 0.9997; \emph{m} = 1.003). Indeed the mechanism effectively acts upon empirical firing rate. }
	\label{Fig4}
\end{figure}

The goal of MANA's signature mechanism was to self organize the target firing rates in a SORN-like model so as to reproduce the roughly lognormal distribution of firing rates which has been consistently reported in the literature in both spontaneous and evoked activity \cite{hromadka2008sparse, mizuseki2013preconfigured, o2010neural, hirase2001firing}(see \cite{buzsaki2014log} for a review). To that end (as it is the foundation of many subsequent results) the ability of the model's formalisms (borrowed from the Log-STDP literature \cite{gilson2011stability}) to produce the desired roughly lognormal distribution of TFRs is of primary concern. This was indeed the case across (and within) 40 networks of 924 neurons each(See. \ref{Fig3} A and \ref{Fig4} A). However, researchers cannot directly measure a neuron's TFR (only the mean firing rate over some time interval) and thus the distributions of firing rates reported in the literature are empirically observed averages of \# of spikes/some time interval. Therefore it was necessary to check that MANA's empirically observed firing rates were also roughly lognormal and tracked well with the TFRs, the latter being necessary to validate the effectiveness of the combination of the meta-homeostatic and homeostatic firing rate plasticities. Indeed empirically observed firing rates of the 36,960 neurons across all 40 networks over the last 700s of simulated time were roughly lognormal and could be fitted to their TFRs with $R^2$ = 0.9997 indicating that neurons' empirical average firing rates were very close to their self-organized target values  (See \ref{Fig4} B \& C). 
		
\subsection*{Synaptic efficacy statistics}
\begin{figure}[!h]
	\includegraphics[width=\linewidth]{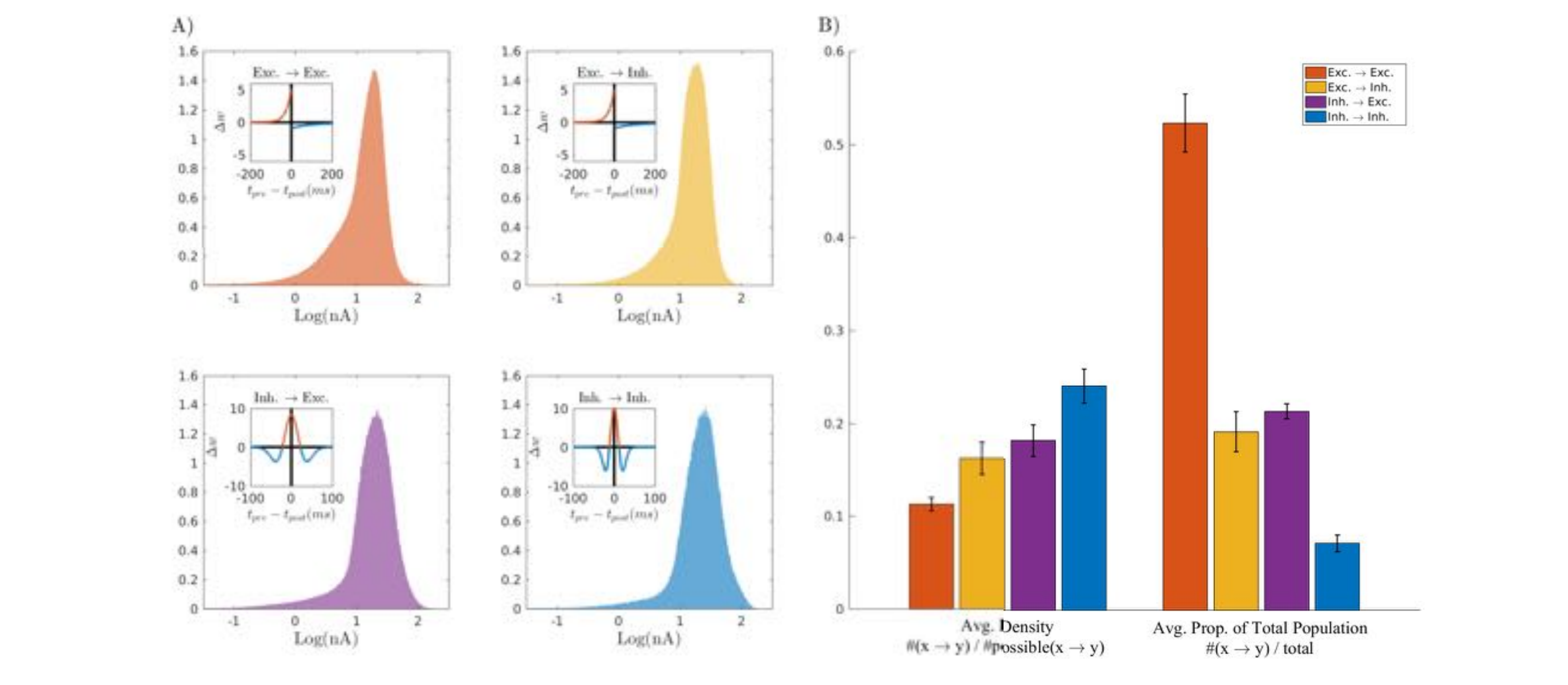}
	\caption{{\bf Heavy-tailed synaptic strengths and relative populations}
		A) Read left to right and top to bottom, histograms of the synaptic strength distributions of each type (Exc.\textrightarrow Exc., Exc.\textrightarrow Inh., Inh.\textrightarrow Exc., Inh.\textrightarrow Inh.) across all 40 networks. Inset in each is the STDP window for the STDP rule governing synapses with the respective types of source and target neuron. Exc.\textrightarrow Exc. and Exc.\textrightarrow Inh. used Hebbian windows \cite{bi1998synaptic}\cite{fino2008cell}, while Inh.\textrightarrow Exc. and Inh.\textrightarrow Inh. used a ``Mexican hat'' window \cite{woodin2003coincident}. B) Following the same color-coding in (A), the densities and populations of synapses of different types after self-organization. The left gives density in terms of number of synapses of that type that existed divided by the number possible, for a 1000 neuron network with 800(200) excitatory(inhibitory) neurons 639,200 Exc.\textrightarrow Exc. connections are possible and thus the first (orange) bar on the left series shows that on average ~11\% (\~70,000) of these possible connections existed across the 40 networks and so on for EI, IE, and II connections. The right-hand series provides the average number of synapses of a given type as a proportion of the total number of synapses. For instance while II connections have the highest density meaning the greatest number of possible connections actually existed, it is by far the lease prevalent kind of connection because inhibitory neurons make up only 20\% of the total neuronal population. \footnotesize }
	\label{Fig5}
\end{figure}

Synaptic efficacies of all varieties (Exc. \textrightarrow Exc., Exc. \textrightarrow Inh., Inh. \textrightarrow Exc., and Inh. \textrightarrow Inh.) also took on heavy tailed distributions (see Fig.\ref{Fig5}). This represents the first result which was not in some way intrinsically built into MANA. Heavy-tailed distributions of synaptic efficacy have been found in SORN models among the network's excitatory synapses \cite{miner2016plasticity}, however here we are able to produce such a distribution among MANA's inhibitory synapses as well due to our inclusion of iSTDP. Interestingly the distributions of synaptic efficacies for inhibitory neurons (Fig. \ref{Fig5} C \& D) were quite similar in shape to the excitatory synaptic efficacy distributions (Fig. \ref{Fig5} A \& B)), despite the former having very different STDP windows from the latter.  To the author's knowledge MANA with iSTDP is the first complete network model to approximately produce or otherwise self organize the heavy-tailed distribution of \emph{inhibitory} synaptic currents found in the literature \cite{borst1994large, brussaard1997plasticity, nusser1997differences}. In general Log-normal distributions of synaptic efficacy have been found in both living tissue \cite{song2005highly,lefort2009excitatory, feldmeyer2002synaptic} and in functional connectivity \cite{nigam2016} (again, for a review see \cite{buzsaki2014log}). The distributions of synaptic efficacy here and in SORN models \cite{miner2016plasticity} seems to carry with it a roughly lognormal shape but with a much heavier left-hand tail. Given the strong possibility of under-sampling of very weak synaptic connections in empirical studies the distributions here would seem at the very least plausible. 
	
\subsection*{MANA reproduces wiring differences between high and low firing rate neurons}
\label{Sec.MANA_FR_wire}
\begin{figure}[!h]
	\includegraphics[width=.8\linewidth]{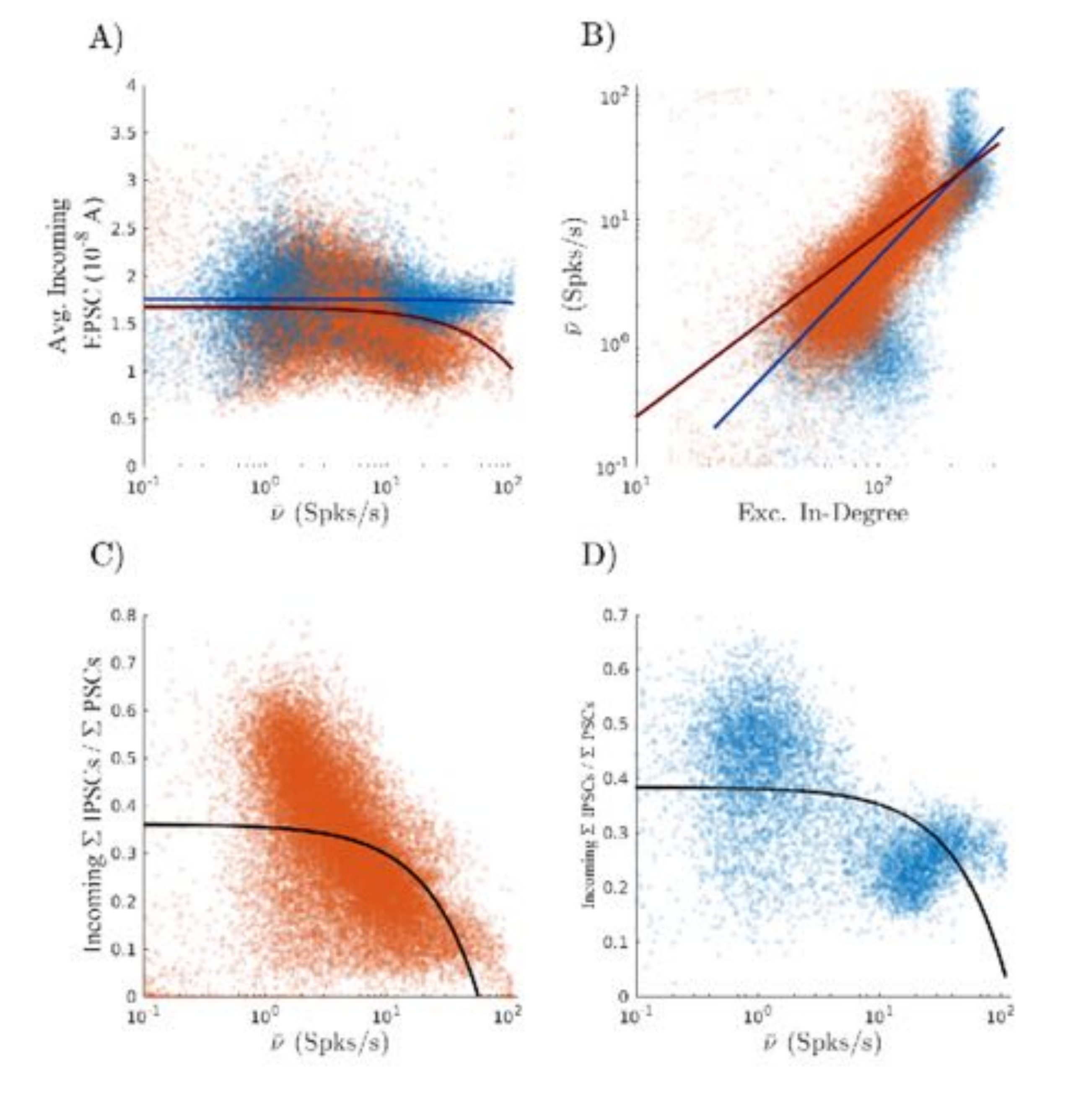}
	\caption{{\bf Qualities of low and high firing rate neurons}
		   Across all 40 networks; orange(blue) represents excitatory(inhibitory) neurons: A) Average strength of incoming excitatory synapses plotted against firing rate, demonstrating that high firing rate neurons do not receive stronger incoming excitatory connections than low firing rate neurons. B) Excitatory in-degree is a good predictor of firing rate. Combined with (A) this demonstrates that high firing rate neurons receive \emph{more, not stronger} incoming excitatory connections than their low firing rate counterparts as found in \cite{benedetti2012differential}. C\&D) The proportion of total incoming drive which is inhibitory for each neuron plotted against TFR. Here we see that low firing rate neurons of both types receive more inhibition while high firing rate neurons regardless of type receive less, a relationship known to exist in the data \cite{benedetti2012differential}\cite{yassin2010embedded}. }
	\label{Fig6}
\end{figure}

\paragraph{}
Studies performed on transgenic mice expressing a green fluorescent protein which was coupled to the activity-dependent c-fos gene demonstrated key differences in the wiring of between neurons which expressed c-fos (were more active) and did not (had a history of less activity) \cite{barth2004alteration}\cite{yassin2010embedded}\cite{benedetti2012differential}. If MANA's self-organization scheme (which includes the emergence of analogs to the c-fos expressing, highly active neurons in the data in the form of neurons with high TFRs) is plausible, then ought to be expected that differences in wiring similar to those found in \cite{benedetti2012differential}, would be observed. 
\paragraph{}
Indeed MANA was able to replicate many observed differences in wiring from \cite{yassin2010embedded} and \cite{benedetti2012differential}. These include: 1) c-fos expressing (more active/high firing rate) neurons have more afferent excitatory connections, 2) that the mean uEPSPs of those connections were not stronger than the mean excitatory connections impinging on neurons which did not express c-fos (i.e. high activity neurons have \emph{more} but not stronger incoming connections), and 3) Excitatory neurons expressing c-fos received decreased inhibition compared to their less active counterparts. Across all 40 networks both (1) and (2) were clearly the case with MANA (see Fig. \ref{Fig6}). This behavior was not programmed into the network. Synaptic normalization did indeed operate so as to give higher firing rate neurons more total allowed incoming current, however this did not guarantee that the settled upon total incoming current would come from large numbers of synapses with similar strengths to those of lower firing rate neurons as opposed to smaller numbers of much stronger excitatory synapses. Additionally, while neurons were able to manipulate their total incoming Exc./Inh. ratios, no mechanism was preprogrammed in a manner which forced high TFR neurons to have reduced inhibitory drive. As outlined in \cite{Luz2012} it may be the case the iSTDP is exerting negative feedback  on the higher firing rate neurons. It is also possible that these neurons are simply accumulating more excitatory synaptic connections.

\subsubsection*{MANA without the ``M'': MHP Specifically Accounts for Features of High Firing Rate Exc. Neurons}
\label{SubSec.MHP_FR_wire}

Of the mechanisms which together comprise MANA, meta-homeostatic plasticity is the most innovative, speculative, and central to the significance of this work. Similarly sophisticated versions of the SORN have studied in detail the contributions of Homeostatic Plasticity, Synaptic Normalization, Short-Term Plasticity,  STDP, and structural plasticity to a model which includes all of them \cite{miner2016plasticity}. Further, certain mechanism's modes of action simply operate in such a way that their lesioning would have obvious repercussions on the network: e.g. without STDP there would be no rule governing growth or pruning thus removing any self-organized topological features, without HP neurons would be severely hampered in their ability to maintain their target firing rates, and so on. This sort of catastrophic collapse of key network function(s) often obscures the precise role of the lesioned mechanism in the emergence of different phenomena, and one must also be careful to consider the implicit assumption that a \emph{one-to-one} mapping between a given component/mechanism and a resulting feature can be reasonably assigned. By analogy, consider that damage to the reticular formation (RF) would almost certainly result in immediate loss of consciousness in the organism (via death), but this is not taken as evidence that the RF is responsible for or otherwise the seat of consciousness. Considering on all these factors: the importance of MHP to the contribution of the model, the fact that other investigators have scrutinized lesions of other mechanisms, and that lesioning MHP would not result in an obvious catastrophic failure, we chose to focus on the results of lesioning MHP specifically so as to ascertain what phenomena it in conjunction with the other mechanisms is responsible for.

In order to focus as much as possible on MHP specifically, a network was simulated wherein only MHP was inactive. Instead the target firing rates of all other neurons were assigned at the beginning of the simulation and frozen from that point forward. TFRs were drawn randomly from a lognormal distribution with the same scale and location parameters as the best-fit lognormal distribution for the TFRs of the in-tact network to which the lesioned network was being compared. Further, all synaptic normalization sums were selected accordingly as well according to equation \ref{Eqn:16}, and likewise normalization did not activate until total incoming synaptic strength reached that value via STDP. Otherwise all other mechanisms operated in exactly the same manner. 

\begin{figure}[!h]
	\centering
	\includegraphics[width=\linewidth]{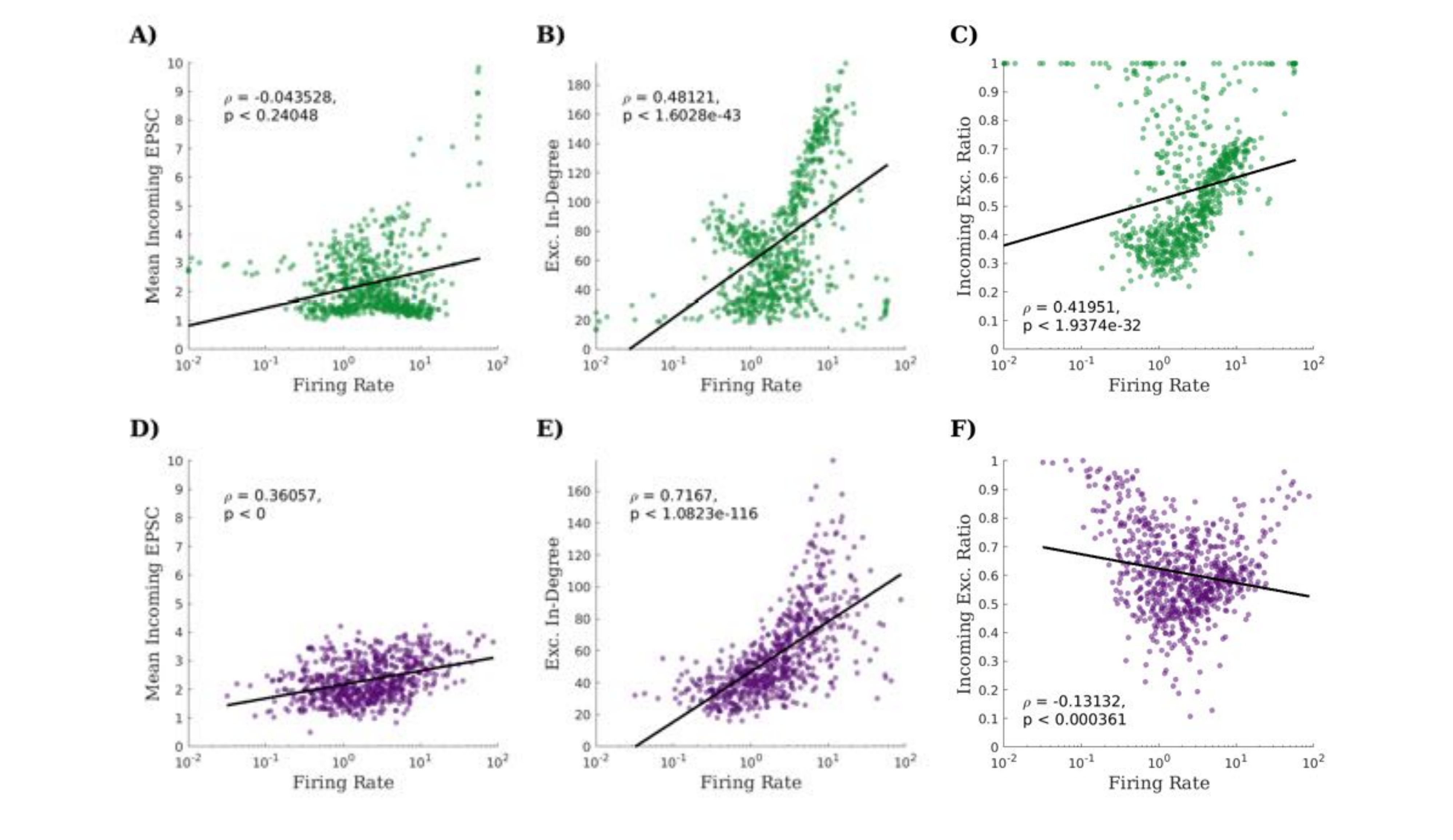}
	\caption{{\bf MHP crucial to relationships between firing rate and topology}
		\textbf{(A-C): In-tact MANA: }Here the relationships found in \cite{yassin2010embedded, benedetti2012differential} all hold: firing rate is not correlated with average strength of incoming excitatory synapses, but is correlated with both excitatory in-degree and the ratio of excitation to inhibition (higher firing rate neurons receive less inhibition); (D-F): MHP lesioned network: As opposed to what has been found in the data, here average incoming excitatory synaptic strength is tightly correlated with firing rate and higher firing rate neurons do not receive significantly less inhibition than their less frequently firing counterparts. \textbf{(ALL)} Lines are best least-squares linear fits with $log_{10}$ of the firing rates projected back onto the logarithmic x-axis. Rho and p-values were calculated using Spearmann correlation.  \footnotesize }
	\label{Fig7}
\end{figure}

MHP was generally found to be crucial for several key organizational features of the network topology and in particular how synaptic topology related to firing rate. For instance, as shown in Fig. \ref{Fig7}, without MHP the network is unable to replicate the specific features of high firing rate cells found in living tissue in \cite{yassin2010embedded} and \cite{benedetti2012differential}. When TFRs were not allowed to self-organize higher firing rate excitatory neurons possessed both higher excitatory in-degree and higher average incoming excitatory synaptic strength. Further there was not a clear relationship whereby higher firing rate excitatory neurons received less inhibition. In both the no-MHP and in-tact case synaptic normalization was in effect meaning that the total excitatory sums for neurons with similar firing rates between the two is similar. The implication of an increasing mean with increasing in-degree and a similar total constant implies that for the no-MHP case synaptic connectivity was less equitable, that is the synaptic connections from some neurons must--in the no-MHP case--be much stronger than others in order to have a higher mean all other things being equal. This further implies. given the causal nature of STDP, that in the no-MHP case pre-synaptic neurons had more varied causal relationships with their post-synaptic partners, possibly indicating a higher degree of structure or selectivity with respect to the qualities of pre-synaptic neurons (given the more uniform causal interactions) on the part of post synaptic neurons when TFRs are allowed to self-organize according to MHP. 

\begin{figure}[!h]
 	\centering
 	\includegraphics[width=\linewidth]{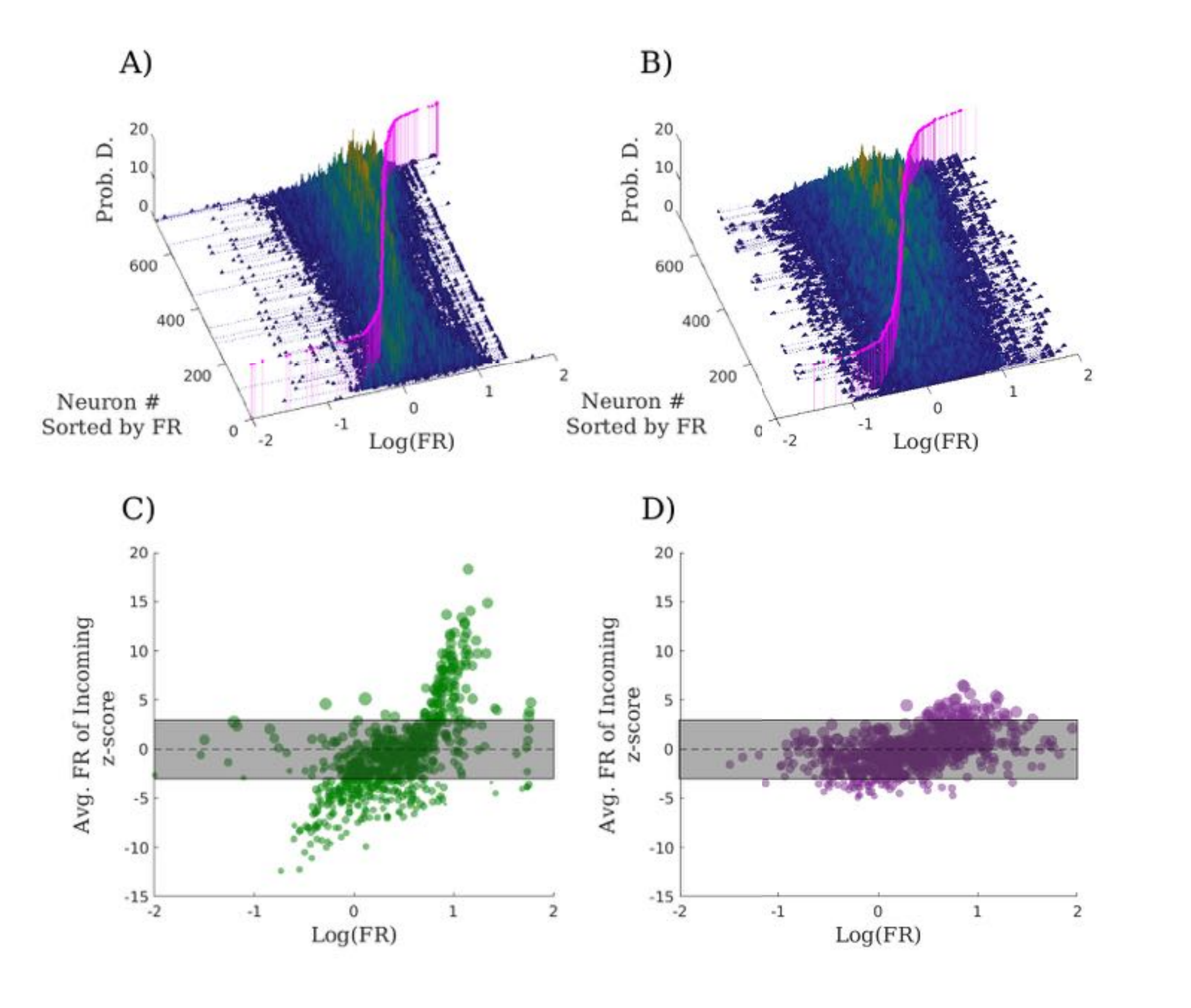}
 	\caption{{\bf MHP leads to preferential attachment based on firing rate}
 		A) A series of histograms of the firing rates of excitatory pre-synaptic neurons for each excitatory neuron in an in-tact network. Histograms are side-by-side and in ascending order (y-axis) based on the firing rate of the post-synaptic neuron. The x-axis corresponds to the histogram-bins for the firing rates of pre-synaptic neurons. The magenta dots represent the firing rates of the post-synaptic cells and are placed above their respective histograms for visibility. Lines have been drawn directly downward from each point to show where each neuron's firing rate is positioned relative to the firing rate distribution of its in-neighbors. Notice the ridge in the distributions which straddles the firing rates of the post-synaptic cells. B) Same as (A) except for the no-MHP network. C) Each dot represents the average firing rate of each (exc.) neuron's (exc.) pre-synaptic neighbors in the in-tact model and is displayed as a z-score relative to the distribution of firing rates of (exc.) pre-synaptic neurons for that neuron across 1000 null models (degree-preserving rewires). Neurons which fire below the mean tend to select much less frequently firing neurons than would be expected by chance, while those that fire above the mean receive connections from neurons with much higher firing rates than would be expected by chance. The transparent gray box represents $\pm$3 standard deviations and in this context denotes values not significantly different from random chance. D) Same as (C) except with the no-MHP network. \footnotesize }
 	\label{Fig8}
 \end{figure}

Since neurons with similar firing rates is one way to possess similar causal interactions with a target cell, the firing rates of pre-synaptic cells relative to post synaptic cells was investigated. To ascertain if MHP specifically was responsible for some higher degree of selectivity or topological organization with respect to high firing rate neurons, the average firing rates of pre-synaptic neurons to each neuron in both networks in both cases was investigated. Specifically, when histograms of the average firing rate of pre-synaptic cells are placed side by side and ordered by average firing rate of the post-synaptic cell an interesting trend emerges. In the no-MHP case distributions of pre-synaptic firing rates (despite the trend of their means upward) appear very similar in shape and indeed aside from the increasing means the overall shapes and locations of the distributions appear very similar with increasing firing rate. That is to say that in the no-MHP case excitatory neurons do not appear to be selecting for their pre-synaptic neighbors' firing rates nonrandomly. In fact the increasing mean, though ostensibly a sign of differences in the selectivity of pre-synaptic firing rate with post-synatic firing rate could come about by chance. The larger the sample size taken from a heavy-tailed distribution like the lognormal distribution, the more likely the sample is to contain members of the tail which will skew the mean upward converging on the actual mean of the distribution. In order to test this the average firing rate of the pre-synaptic cells in each network was measured and averaged across 1000 null models in the form of degree-preserving rewires. 
 In Fig. \ref{Fig8} we can see the results, namely that the average firing rates of pre-synaptic cells for the no-MHP network are for the most part not significantly different than chance, while for the in-tact network less (more) frequently firing neurons appear to receive connections from much less (more) frequently firing neurons than would be expected by chance (in some cases by nearly 20 standard deviations). Similarly in the in-tact case we observe that the post-synaptic firing rate ordered pre-synaptic firing rate histograms reveal a ridge which straddles the post-synaptic firing rates initially from above then crossing over to above. Other than in places where the post-synaptic firing rate is near the network mean the distribution of pre-synaptic firing rates appears somewhat more homogeneous, collecting near the post-synaptic firing rates. In other words, MHP appears to result in a network whereby higher firing rate neurons appear to select incoming neighbors with firing rates directly below their own, and much higher than the network mean. Similarly, infrequently firing neurons select incoming neighbors with firing rates directly above their own, which for many is much below the network mean. It would seem that this configuration of ``inside-out'' firing-rate-preferential connectivity provides at least one account of how high firing rate neurons might possess more, but not stronger incoming excitatory connections. The fact that neurons in the network with MHP tended to receive more connections from neurons close (directly above/below depending upon what side of the TFR mean the neuron's TFR is) to their own firing rate would appear to also agree with the observation in \cite{yassin2010embedded} that neurons expressing the activity-dependent c-fos gene (which fired more frequently and thus had more similar firing rates) were more likely to connect with one another, further bolstering MHP as a mechanism which can account for the topological and organizational features of high firing rate neurons. 
  
This configuration likely comes about as a direct result of the repulsive force exerted on post-synaptic TFRs from the difference in pre- and post-synaptic EFRs. For instance: In order to maintain a high TFR a neuron must have a significant number of neighbors whose EFRs are consistently below its own. In particular, since the force exerted on TFR drops off exponentially with distance in EFR-space, high TFR neurons actually require many of their incoming neighbors to possess lower, but still very similar EFRs to their own. This would on the surface appear to be reason enough for this configuration since the selection process for possessing a given TFR requires nearby neurons in EFR-space closer to the network mean to support a neuron's TFR at its location in TFR space. However, this does not mean that these neurons cannot make or would necessarily not maintain connections to neurons of all different firing rates in the same way as post-synaptic neurons in the no-MHP network do. Rather this configuration and MHP being necessary to it either appears as a result of MHP acting collectively on all neurons together and/or that neurons with similar firing rates are more likely to be (or have more opportunities to be )causally linked to the post-synaptic cell, thus out-competing others. However, if the latter were solely responsible we'd expect to see this configuration in the no-MHP network.

\subsection*{Synaptic Topology}

\subsubsection*{MANA produces nonrandom topological features}

\begin{figure}[!h]
	\centering
	\includegraphics[width=\linewidth]{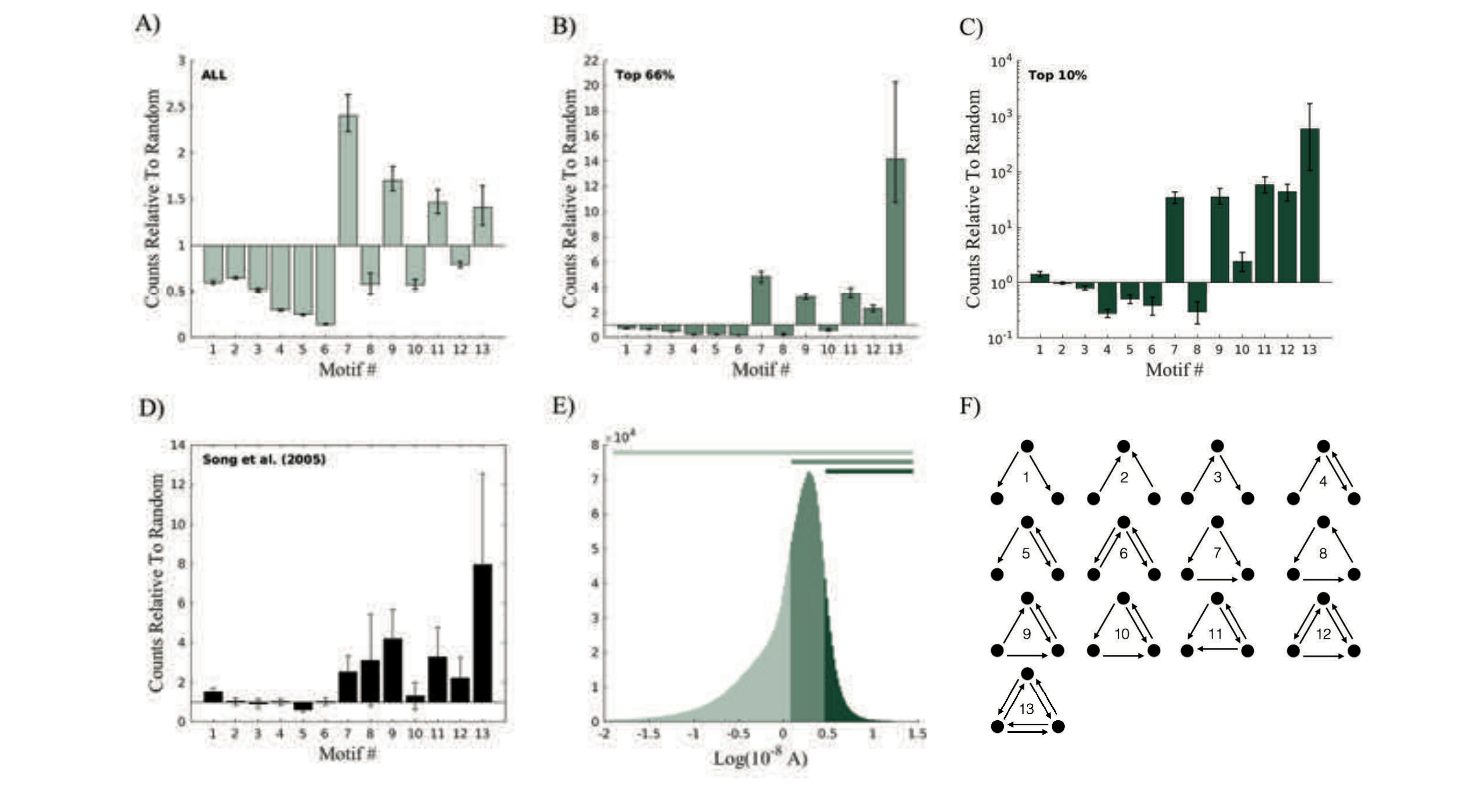}
	\caption{{\bf Distribution of Triadic Motifs across 40 networks}
		The distributions of the statistical over-representations of different triads within the Exc. \textrightarrow Exc. subnetworks of all 40 networks with different thresholds. A) Motif over-representation when all synapses are included. However weak synapses are more easily not detected empirically. B) Motif over-representations when only the top 2/3rds of synapses by strength are considered. C) Same as (A \&B) when only the top 10\% of synapses are considered, \emph{notice the shift to a log-scale} due to the extreme over-representations at this threshold. Notice that the over-representations generally (and especially with motifs involving bidirectional connections) become stronger when only stronger synapses are considered, consistent with \cite{song2005highly} D) The triadic motif over-representations reported in \cite{song2005highly} found between layer V tufted neurons in rat visual cortex. E) A histogram of all Exc.\textrightarrow Exc. synapses color-coded to demonstrate where in the overarching distribution of synaptic strengths (A-C) are sampled from. F) The motif key. \footnotesize }
	\label{Fig9}
\end{figure}

Patch-clamp studies of connectivity by Perin et al.\cite{perin2011synaptic} and Song et al. \cite{song2005highly} have shown that excitatory neurons cluster in non-random patterns. Interestingly this result has also been found in studies using effective/functional connectivity \cite{shimono2015functional}. Using the same methods as in \cite{song2005highly} (for comparison purposes), whereby null models were derived from base connection probability, the over-representation of specific 3-motifs was examined (Fig. \ref{Fig9}). Not only did certain 3-motifs appear overrepresented in approximately the same way, but this over-expression became exceedingly more prevalent when only stronger synapses were considered, replicating the findings of \cite{song2005highly}, whereby the over-represented motifs were comprised of stronger synapses, thus forming a network backbone of nonrandom triadic connections.

Using similar methods to those developed in \cite{perin2011synaptic} the over-representation of higher numbers of connections within 3, 4, 5, and 6 neuron clusters was examined here, so as to compare between the living data and MANA. Given the size of the excitatory subnetworks (typically ~780 neurons) and the number of network data sets (40), slight deviations from the techniques in \cite{perin2011synaptic} were required. In order to be statistically rigorous while maintaining computational tractability the following scheme was used: As in \cite{perin2011synaptic} our null models consisted of a random-rewire of the network which preserved node degree (in and out) as well as the number bidirectional and unidirectional connections. For each cluster size 10 million \emph{distinct} combinations of 3-6 nodes were randomly chosen from each of the 40 networks and the number of connections found within each cluster was recorded. For each of the 40 networks 1000 null models were generated and 10,000 unique combinations of 3-6 neurons were sampled. This generated a distribution of the quantity of connections within each cluster type for each of the 40 networks as well as a null distribution for each of the 40 networks. Despite the discrepancy results remain comparable since the same null models were used. The only difference exists in the sampling since a combinatorial explosion makes a full survey of the space of null models impossible. A 2-way KS-statistic between the distributions between the null models and the 40 networks for each number of connections in each cluster was used to calculate our p-values.

\begin{figure}[!h]
	\centering
	\includegraphics[width=.8\linewidth]{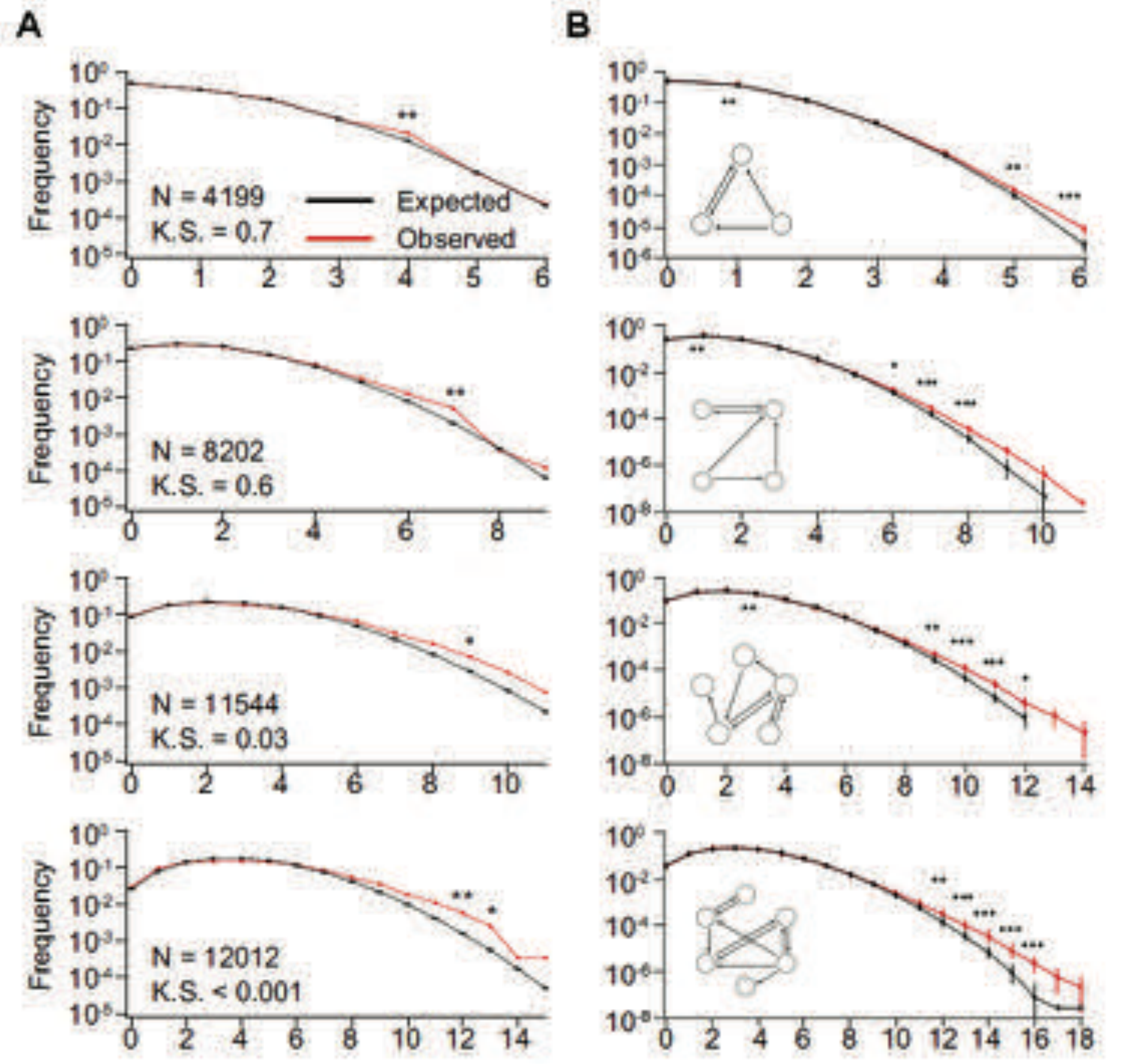}
	\caption{{\bf The over-representation of highly connected groups}
		A) The comparison of observed (red) and expected (black) frequencies of number of synaptic connections in clusters of size 3, 4, and 5 taken directly from \cite{perin2011synaptic}, which derived these values from recordings of up to 12 layer V pyramidal neurons taken from slices of rat somatosensory cortex. B) A comparison of the number of observed vs. expected prevalence of numbers of synaptic connections in different clusters as in (A) across the Exc. \textrightarrow Exc. subnetwork of all 40 simulated MANA networks.}
	\label{Fig10}
\end{figure}

The results of this analysis can be seen in Fig. \ref{Fig10}, which broadly speaking demonstrates that the model self-organizes more tightly coupled clusters than would be expected by chance, as has been found in patch-clamp \cite{perin2011synaptic} and effective connectivity \cite{shimono2015functional} studies.

\subsubsection*{MANA self-organizes specialized groups}

The laminar structure of cortex is a well studied phenomena in neuroscience. Different layers of mammalian cortex are populated by different cell types and have particular relationships to one another. In particular laminar layers differ with respect to where their inputs originate, where their outputs target, and the degree to which they serve as inputs ans/or outputs to the column as a whole. For instance layer IV is known to receive significant amounts of input from thalamus ``core'' or C-type cells \cite{jones1998viewpoint} and send a great deal of outputs to layers II/III \cite{lefort2009excitatory, yassin2010embedded, benedetti2012differential, harris2015neocortical}. This thalamus \textrightarrow Layer IV \textrightarrow Layers II/III pathway has been studied extensively as being central to early cortical processing of inputs--particularly in barrel cortex \cite{armstrong1987spatiotemporal, armstrong1992flow, ahissar2000transformation, brecht2002dynamic, lefort2009excitatory}. Reconstructions of the connectivity between cortical layers in barrel cortex demonstrate that the layers differ greatly in where within the column they send and receive synaptic connections and the degree to which they connect to themselves \cite{lefort2009excitatory}. Computer simulations of this reconstruction demonstrated that stimulation of Layer IV had the greatest chance of spreading activity across the entire column. Indeed certain hodological themes have been identified across species and areas of cortex which have distinct groups of cells whose connectivity implies distinct input/output/recurrent processing roles \cite{harris2015neocortical}.

\begin{figure}
	\centering
	\includegraphics[width=.8\linewidth]{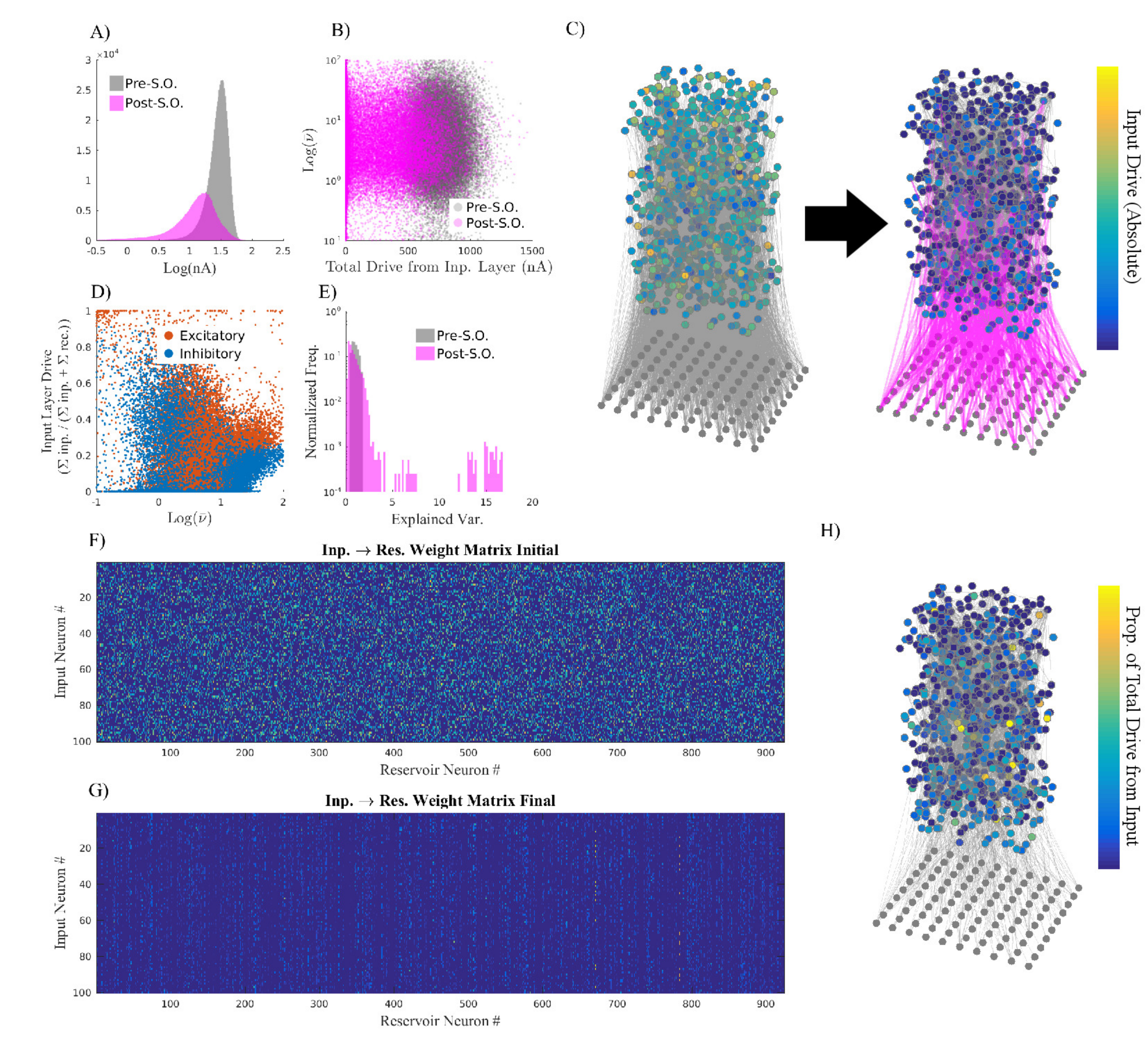}
	\caption{{\bf Diversification of input selectivity}
		A) Overlayed histograms of the synaptic strengths of synapses connecting the input layer to the recurrent reservoir before and after self-organization demonstrating an overall decrease in strengths and quantities. B) Total drive from the input layer (as these synapses were initialized to nonzero values) against final TFR demonstrates no discernible correlation between initial drive and final TFR, ruling out the different starting drives to each neuron as a significant bias toward final activity-related outcomes. C) An example network before and after self-organization with each neuron colored according to total input-layer drive they receive. Notice that most input \textrightarrow reservoir synapses are significantly pruned and that absolute input drive decreases for all neurons and that recurrent connections have grown. Only the top 2.5\% and 10\% of synapses for the input \textrightarrow reservoir and reservoir \textrightarrow reservoir are drawn for visibility. D) A plot of ``inputedness'' defined as the proportion of synaptic drive originating in the input layer against TFR. Neurons exhibit a diversity of levels of inputedness while others settle such that all or nearly all their connections from the input layer are pruned. Some neurons thus receive from the input and project to the rest of the reservoir while others receive only from the reservoir. For inhibitory neurons this indicates the presence of both feedforward and feedback inhibition. E) The explained variances when PCA is performed on the columns of the input \textrightarrow reservoir weight matrices. Fewer principal components are required to explain more of how reservoir neurons select from the input. F\&G) An example input \textrightarrow reservoir weight matrix before and after self-organization demonstrating the evolution from a randomized 25\% dense matrix into a matrix with blatant organization and specifically input correlations to reservoir neurons (noted as a prerequisite for lognormal firing rates \cite{koulakov2009correlated}) H) An example network where each neuron is colored by proportion of total drive from the input layer. Notice significant diversifications as well as a significant number of neurons which have lost all input-layer drive.    }
	\label{Fig11}
\end{figure}

The recurrent MANA reservoir is driven by 100 input neurons which are completely controlled by the experimenter, receive no inputs from the MANA reservoir, and otherwise lack any sort of autonomous dynamics. This ``input layer'' is not part of the network proper, though the synapses connecting the input to the reservoir are. Consider that at an initial input density of 25\% each reservoir neuron was on average connected to 25 input neurons, considering that these connections were random this means the number of incoming connections would take the form of a binomial distribution with p=0.25 of success and n=100 trials. Meta-homeostatic plasticity ensures directly that neurons have different preferred levels of activity and, in accordance with the TFR, different incoming innervation. However MHP in no way dictates or biases the neuron with respect to \emph{where that innervation comes from} . If no functional distinctions were occurring in MANA with respect to the input layer/signal, then it stands to reason that the degree to which neurons receive input from the input layer would not change significantly by the end of the simulation and/or that each neuron would receive roughly the same amount of innervation from the input layer. This was not the case. Neurons in the MANA reservoir took on a wide variety of different levels of innervation from the input layer, in particular a large proportion of neurons lose all input layer innervation, becoming fully recurrent, which distinguishes them from neurons which retained significant input layer drive (see Fig. \ref{Fig11} D-H). Inputs from the input layer to the neurons which retained their input layer drive were also correlated as seen in Fig. \ref{Fig11} G. The implication of this configuration is that MANA reservoir self-organize such that there exist specific neurons which handle external drive, while others do not. External signals must first pass through these neurons with significant (correlated) input layer drive, before coming into contact with the other neurons in the network. 

Notably, initial innervation from the input layer is a poor predictor of TFR, indicating that the latter is determined by properties of the input \emph{patterns} and the self-organization of the MANA layer much more so than by initial conditions (see Fig. \ref{Fig11} B). Thus the very small amount of initialization in MANA (i.e. the weights from the input-layer to the reservoir) did \emph{NOT ultimately bias reservoir TFRs}. 

\begin{figure}[!h]
	\centering
	\includegraphics[width=0.35\linewidth]{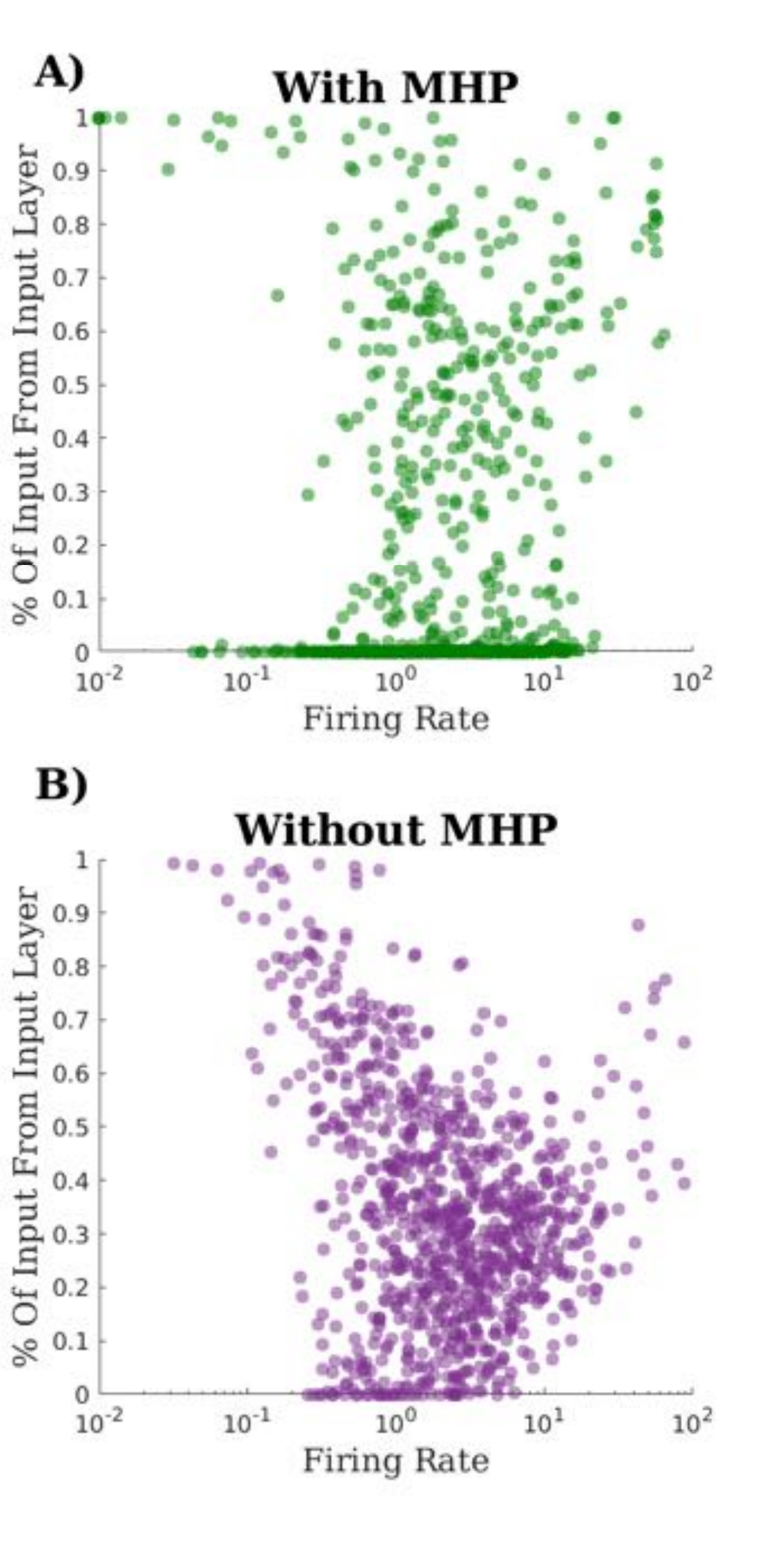}
	\caption{{\bf Input-Selectivity reduced without MHP}
		\footnotesize A)  Firing rate plotted against the amount of excitatory drive from the external input layer for a single in-tact MANA network using MHP. Notable here is the fact that a majority of the neurons receive literally no input-layer drive meaning that all external input must first pass through specific members of the recurrent reservoir before they can affect other reservoir neurons. B) Same as (A) except for a network with no MHP where TFRs were initialized to values drawn from a lognormal distribution. While there is significant diversity with respect to input-layer drive, a distinct segregation between neurons which do and do not receive external drive is not present.      }
	\label{Fig12}
\end{figure}

\begin{figure}[!h]
	\centering
	\includegraphics[width=.8\linewidth]{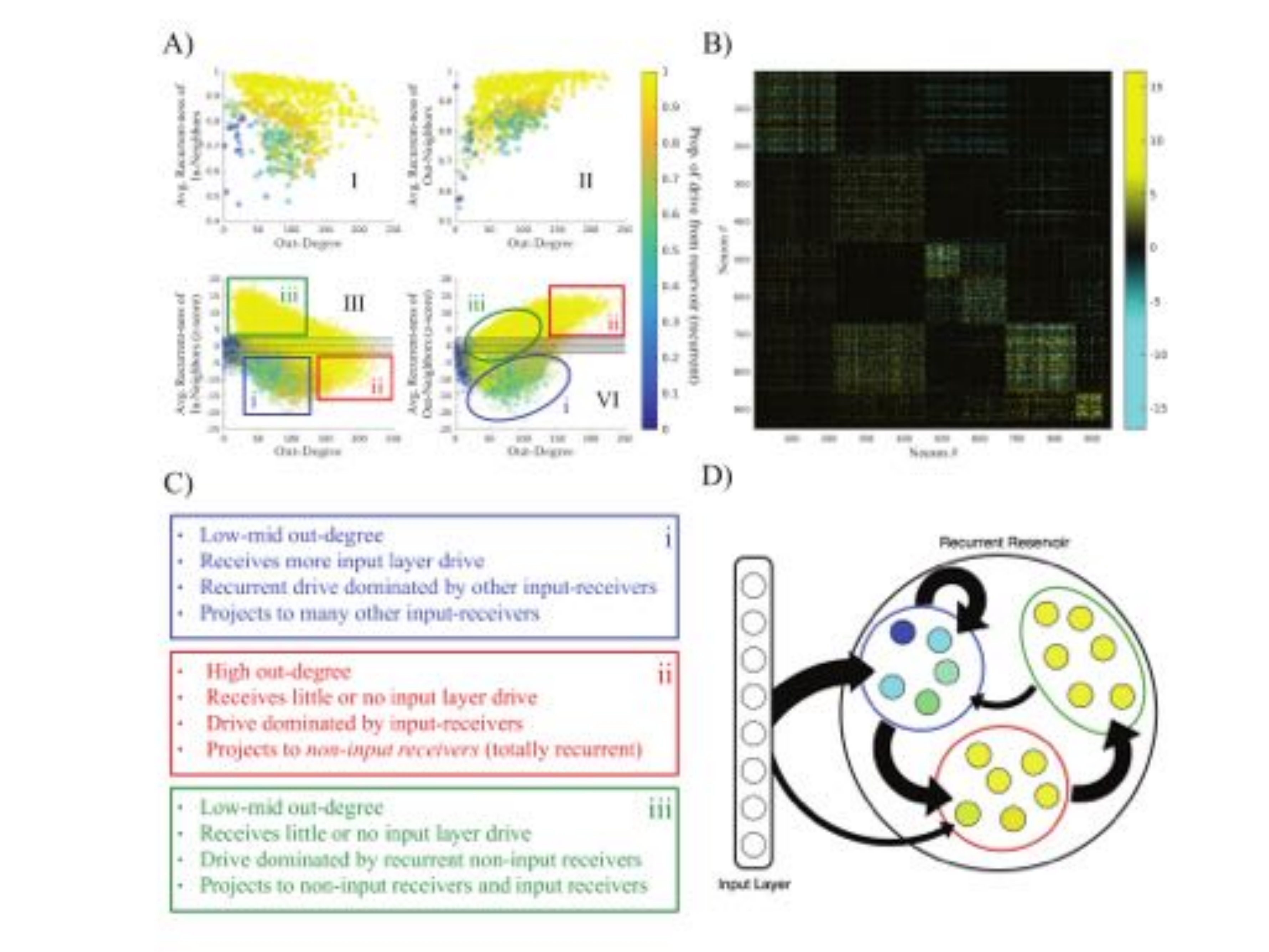}
	\caption{{\bf Emergence of Layered Structure/Diversification of Recurrent Selectivity}
		\footnotesize A.I) Scattering of the out degree of each neuron (x-axis) against the average proportion of incoming drive received from the recurrent layer (vs. the input layer) of each neuron's incoming neighbors for a single representative Exc.\textrightarrow Exc. subnetwork. Points are colored by the proportion of recurrent drive (``recurrentn-ess'') possessed by each neuron themselves. Notice that there is a group of neurons with high out-degree which are yellow and thus themselves have high recurrent-ness, but which receive input from many more neurons with low recurrent-ness. A.II) The same as (A.I) except averaging over outgoing neighbors. Notice that the same group of high out-degree, high recurrent-ness neurons from (A.I) send projections to neurons of high recurrent-ness. A.III \& A.IV The same as (A.I)\& (A.II) except across all 40 Exc.\textrightarrow Exc. subnetworks and normalized by z-score against their respective degree-preserving null models. All points outside the transparent gray box correspond to z-scores with p \textless 0.01. Here 3 groups of interest are highlighted. B) An example reservoir \textrightarrow reservoir weight matrix sorted using the OSLOM community detection algorithm\cite{lancichinetti2011finding}, demonstrating significant modular organization post-self organization. C) Descriptions of the highlighted portions of (A.III) \& (A.IV). D) A hypothetical/idealized example network architecture following from (A) whereby a group of neurons receives significant input drive and neurons which receive exclusively recurrent drive are divided into two groups, one of which receives drive from the neurons which are driven by the input layer and another which receives only from other reservoir neurons.   }
	\label{Fig13}
\end{figure}

In addition to the amount by which synaptic inputs to each neuron from the external input layer changed, the proportion of each neuron's input which originated in the input layer was considered. This gives a more subtle quantitative measure of the neuron's role in the network with respect to the input layer. To this end, each neuron is assigned an ``inputtedness'' score, which is 0 if a neuron only receives inputs from other neurons in the recurrent reservoir and 1 if a neuron receives all its synaptic inputs from the external input layer. This reveals that MANA self-organizes both feed-forward and feedback inhibition and that these roles are taken on by \emph{different} inhibitory neurons, since some inhibitory neurons have a high inputedness score while others have a score of 0 (Fig. \ref{Fig11}) among other things. Fig. \ref{Fig11} shows that a very large fraction of neurons end the simulation with 0 input from the input layer and are thus fully recurrent. Those that do not receive strong input correlations, which can be seen in the vertical striping of the weight matrix connecting the input layer to the reservoir (Fig. \ref{Fig11}G). It's worth noting that input correlations (which can be seen in the reservoir as well in Fig. \ref{Fig13}B) have been shown to be a prerequisite for lognormal firing rate distributions in neural networks \cite{koulakov2009correlated}. Interestingly this degree of specificity whereby some reservoir neurons cut themselves off completely from the external input layer (thus forcing external input through very specific reservoir neurons before becoming the input to others), was present to a significantly \emph{lesser} degree when TFRs were not allowed to self-organize as shown in Fig. \ref{Fig12} where a single in-tact network is shown in comparison for visibility (as opposed to Fig. \ref{Fig11} D where the same figure is shown across all 40 networks and the number of points obscures the larger number of reservoir neurons with zero or near zero input-layer drive). That is, with MHP while among the neurons which receive some external input-layer drive there is significant variation in the amount which they receive there is a clear distinction between those neurons which do and those which do not receive any external drive at all. This strongly suggests a division of labor which simply does not appear to exist in the no-MHP network wherein nearly all neurons receive some external drive which varies rather smoothly from 0-100\% across all the neurons in the network. In this case nearly all neurons receive some degree of external drive, thus sharing in the task of processing external input as opposed to having input layer processing be the specific domain of a concrete subset of reservoir neurons. 

Knowing that some neurons receive more external input while others receive more internal/recurrent input as well as that some neurons have a preference to the degree to which their incoming/outgoing neighbors receive connections from the input layer implies a division of labor. Thus we should expect some neurons which themselves are all or mostly recurrent to receive connections from neurons which receive external drive if such a division of labor is present. To test this we measured the proportion of total incoming drive originating in the recurrent layer for the incoming and outgoing neighbors of each neuron:

\begin{gather}
	r_j \;=\; \frac{1}{N} \sum_{i \in I_R; w_{ij} \neq 0}^N \frac{\sum_{h \in I_R} w_{hi}}{\sum_{h \in I_R} w_{hi} + \sum_{h \in I_i} w_{hi}}\\
	r_i \;=\; \frac{1}{M} \sum_{j \in O_R; w_{ij} \neq 0}^M \frac{\sum_{i \in I_R} w_{ij}}{\sum_{i \in I_R} w_{ij} + \sum_{i \in I_i} w_{ij}} 
\end{gather}
Where $I_R$ and $I_i$ refer to the set of in-neighbors from the recurrent MANA reservoir and the external input layer respectively. $O_R$ is the set of out-neighbors from the recurrent MANA reservoir (note the absence of a $O_I$ which would be out-neighbors to the external input layer since such connections were not permitted i.e. $O_I \equiv \emptyset$). N and M are the set cardinalities of $I_R$ and $O_R$ respectively or the number of in-neighbors to neuron \emph{j} and out-neighbors to neuron \emph{i} in the recurrent reservoir respectively. In this notation \emph{h} is presynaptic to \emph{i}, which is presynaptic to \emph{j}. Neighbors in $I_i$ were not counted since they themselves received no inputs, however weights from neurons in $I_i$ were counted (otherwise a comparison between drive from $I_i$ and $I_R$ would be impossible). Finally this gives us $r_j$ and $r_i$ for each neuron or the average proportion of recurrent drive across the in- and out-neighbors respectively. 

The picture painted by the results in Figs. \ref{Fig11} \& \ref{Fig13}, which considers only the Exc. \textrightarrow Exc. subnetwork, implies a division of labor reminiscent of the hodology discussed in \cite{harris2015neocortical}, whereby certain populations of neurons receive inputs from specific sources and are specialized to that end. In MANA not only do distinct populations of neurons exist which possess and do not possess input drive, but the results in Fig. \ref{Fig13} indicate that the neurons which do not receive direct input layer drive can be further subdivided into those which receive drive from the neurons which receive input layer drive and ones which do not. In other words a portion of the population that does not receive direct drive from the input layer is highly innervated by the population of neurons which do. These neurons then feed other neurons which receive no input layer drive. Notably, the neurons which receive drive from neurons which receive input layer drive and project to neurons which do not possess the highest out-degrees. Indeed nearly all the high out-degree neurons have this property, while neurons in the other two groups have lower out-degrees. This indicates that these populations differ not only in where they receive inputs and send outputs, but also in some of their intrinsic attributes. That is the populations appear to be composed of different kinds of neurons with different qualities insofar as such a thing is expressible with leaky integrate-and-fire point neurons. In sum, each population has distinct, specific sources of input and targets of output which are arranged such that each population takes as input the output of the last. This input selectivity and arrangement combined with differences in attributes like out-degree is highly indicative of specialization, and has emerged entirely through self-organizing mechanisms acting on much lower level aspects of the network.

\subsubsection*{MANA self-organizes hubs}
	
It is well established that neurons are highly heterogeneous in terms of attributes like firing rate and synaptic degree, having more or less incoming/outgoing connections to other neurons and possibly stronger connections to said neurons \cite{barth2004alteration, hromadka2008sparse, mizuseki2013preconfigured, buzsaki2014log, nigam2016, shimono2015functional, bonifazi2009gabaergic}. Furthermore synaptic structure is thought to posses some scale-free or small world attributes, which have been found in studies of functional connectivity \cite{nigam2016, shimono2015functional, bonifazi2009gabaergic}. Such structures are considered ideal for neural circuits since they represent a compromise between wiring cost and efficiency\cite{sporns2011networks, sporns2007identification}, and indeed hub neurons in line with this topology have been found in studies of functional connectivity \cite{nigam2016, shimono2015functional, sporns2007identification, bonifazi2009gabaergic}. 
	
Over the course of its self-organization, the network not only produces high degree hub neurons, but also settles into a state wherein these hubs are more highly connected to one another than would be expected by chance, forming the so-called "rich-club" \cite{colizza2006detecting}. This particular quality of hubs has been observed both in studies of connectivity of the mammalian microconnectome \cite{harriger2012rich}\cite{schroeter2015emergence}\cite{nigam2016}, and directly in the synaptic connectivity of C. Elegans \cite{towlson2013rich}. The notion of a rich club can extend to any parameters of the nodes in the network since it merely measures whether or not neurons rich in the particular quantity of choice connect to one another beyond chance. The rich-club coefficient for directed networks is defined as:
	
$$ \phi(k) \;=\; \frac{E_{>k}}{N_{>k}(N_{>k}-1)} $$
	
Where $k$ is the richness parameter (synaptic degree unless otherwise specified), $E_{>k}$ is the number of directed edges between nodes where the richness parameter is greater than $k$ and $N_{>k}$ is the number of nodes which posses a richness parameter greater than k \cite{colizza2006detecting}\cite{smilkov2010rich}. However because this value tends to monotonically increase for random networks, richness is typically measured with respect to a null model to produce a "normalized rich club coefficient", which shows how much more (or less) the given network's hub nodes connect to each other than what one would expect from chance:
	
$$\phi_{norm} \;=\; \frac{\phi}{\phi_{null}}$$
	
Where $\phi_{null}$ is the mean rich-club coefficient of 100 networks for which synaptic connections have been rewired, preserving degree distribution, but otherwise randomizing the structure, which is consistent with the literature \cite{colizza2006detecting}. 
	
Rich clubs emerged very robustly in the Exc.\textrightarrow Exc. subnetwork appearing when using in-degree, out-degree, and degree as the richness parameter. These appeared both in the in-tact subnetworks and when only the top 10\% of synapses (within each subnetwork) were used. Rich clubs across the whole network appeared to exist, but less reliably so, with a significant portion (though not a majority) of the 40 networks having no out-degree or degree rich clubs when inhibitory neurons are included. Interestingly a strong in-degree rich-club does appear to be present. Results for the full networks where only the top 10\% of synapses (within each network) were considered appear similar to the in tact full networks. Given the high degree of correlation between in-degree and firing rate this result further bolsters the claim that MANA can replicate the firing rate dependent features of living neurons, since it was observed in \cite{yassin2010embedded} that more active neurons (those with similar levels of activity) tended to preferentially connect to one another. 

\begin{figure}[!h]
	\includegraphics[width=.8\linewidth]{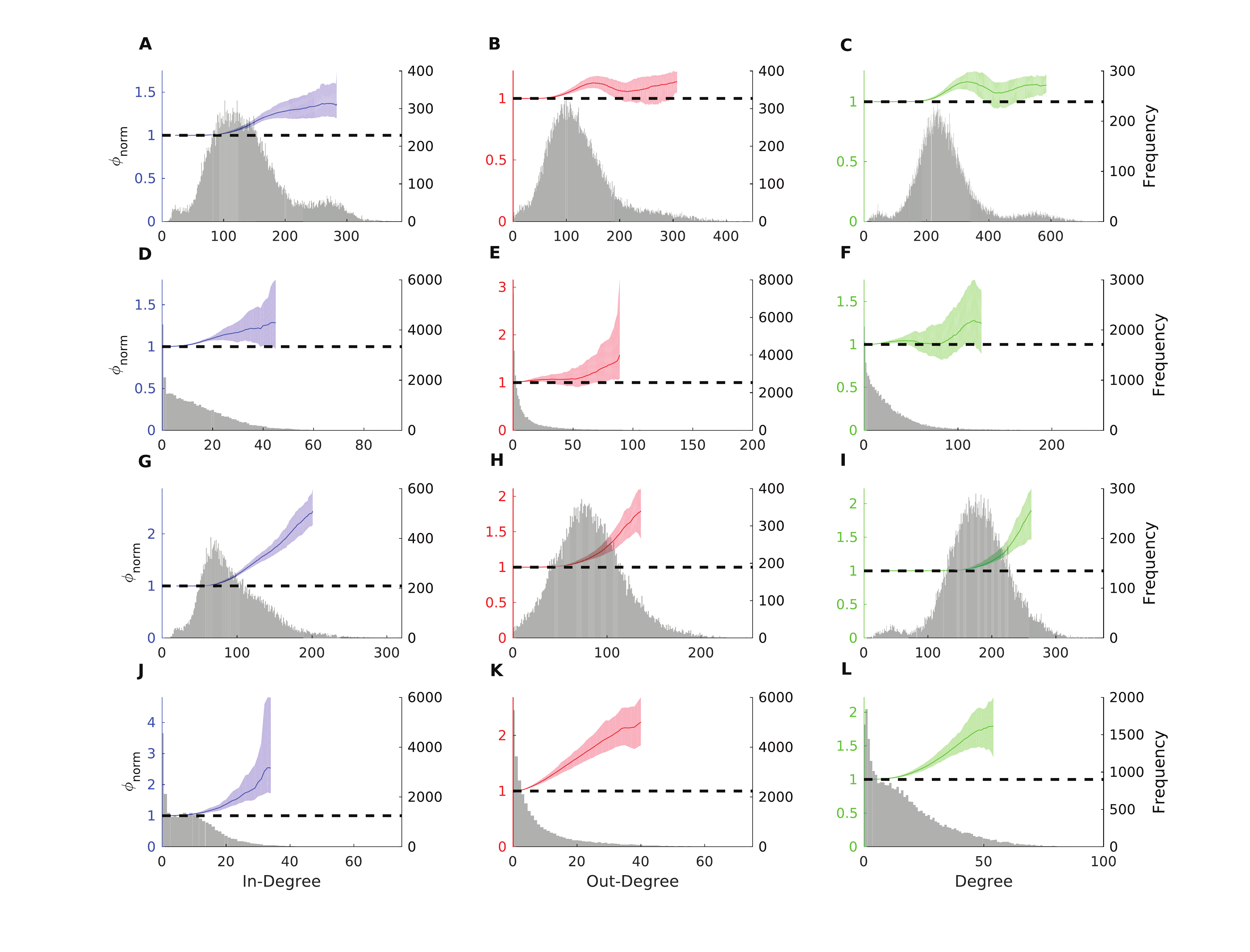}
	\caption{{\bf Rich-Clubs and Degree Distributions}
		\footnotesize (A-C) The in-degree (blue), out-degree (red), and degree (green) average normalized rich-club coefficients $\phi_{norm}$ with standard error represented by the shaded regions across all 40 networks. Gray histograms are the respective degree distributions taken from all neurons across all networks. Average $\phi_{norm}$ is calculated only so long as all 40 networks had at least 2 neurons with the specified degree or higher. (D-F) same as (A-C) except only using the top 10\% of synapses within each network. (G-I) Same as (A-C) except using only the Exc.\textrightarrow Exc. subnetworks. (J-K) Same as (G-I) except using only the top 10\% of Exc.\textrightarrow Exc. synapses.   }
	\label{Fig14}
\end{figure}

Interestingly the model also produces inhibitory hub neurons which have been reported in \cite{bonifazi2009gabaergic} as inhibitory neurons tended to be the highest degree (in and out) in all of the networks. These inhibitory hubs are indeed more connected to other hubs/rich-nodes with respect to in-degree (see Fig. \ref{Fig14} (A)) than would be expected by chance and are thus members of one of the rich-clubs. The implications of a strong in-rich club, but weak or nonexistent out-rich club when inhibitory neurons are present combined with the knowledge that the inhibitory neurons are in fact the richest with respect to those parameters lends itself to the rather interesting conclusion that there must exist two groups of inhibitory neurons ones with high in-degree and ones with high out-degree. Those with high in-degrees are in a rich-club implying that widespread network activity will activate these inhibitory neurons silencing the network, but also themselves, thus allowing for recovery. Those with high out degrees are not in a rich club implying that activation of these interneurons can be used to silence large portions of the network in ways which do not directly inhibit one another. High interconnectivity of inhibitory neurons has been observed in an optogenetic study of acute slices of mouse visual cortex \cite{pfeffer2013inhibition}. In particular, \cite{pfeffer2013inhibition} found that parvalbumin expressing interneurons (the largest type, representing 36\% of the total population) inhibited one another as strongly as they inhibited pyramidal cells. However the 2nd largest group expressing somatostatin heavily inhibited parvalbumin expressing interneurons and pyramidal cells but \emph{did not} inhibit themselves \cite{pfeffer2013inhibition}. In general a high degree of inhibition of inhibition (see Fig.\ref{Fig5}) appears in MANA, but in a highly organized fashion.

%PLOS does not support heading levels beyond the 3rd (no 4th level headings).

\section*{Discussion}

Self-organization is key to how brains develop and change in response to new information. It's been common knowledge for years that no one plasticity mechanism can explain how the characteristic features of cortex arise alone. Models like the SORN have made substantial progress in detailing how some of these mechanisms might interact and why that interaction is beneficial or otherwise evidenced by current data. However, up until now we have lacked a comprehensive model capable of generating a wide array of cortical circuit features from a relative null state, which can serve as the basis for experiments requiring the context of the successful interaction of many mechanisms (as is the case in living tissue). We have demonstrated the first recurrent spiking neural model capable of self-organizing its own TFRs. Beyond that we have demonstrated a large-scale SNN model which self-organizes its inhibitory connectivity both in terms of strength and structure, and combined both these features with other mechanisms so as to ``fully'' self-organize a realistic neural circuit from the ground up. The resulting network is stable in spite of the fact that key aspects (most notably TFRs and connectivity) of the network remain in flux for a substantial portion of its self-organization, showing that stabilizing mechanisms like homeostatic plasticity and synaptic normalization are sufficient to provide stability even in a network with "moving targets" with respect to the target levels of activity for its individual neurons.  
	 
Models including both aspects of neuronal self-organization (differentiation/development and homeostasis) are quite desirable, but typically rare. RNNs have a history of promising a wealth of advantages over feed-forward and more classical AI techniques (and often delivering: \cite{verstraeten2005isolated} \cite{graves2013hybrid}), but are frequently sidelined given their notorious difficulty and impenetrability. Though the reservoir computing paradigm has  Self-organization allows for mechanisms intrinsic to the network to organize around inputs from the problem space, and ideally find an optimal structural and behavioral configuration. Indeed work from \cite{lazar2009sorn},\cite{steil2007online, schrauwen2008improving, ju2013effects}, has shown this to be the case. Although the parameters of the self-organization would likely need some tuning to the application, the point of a self-organizing network, both in the hands of engineers and from the standpoint of a natural system is to reduce what could be a lengthy or expensive parameter search for an optimal network for the problem-domain.

\subsection*{Validity}
	 
The validity of MANA with respect to the aggregate of its parameters and particular design decisions must be understood in the context of the initial goals of the project: design a cortical model which possesses as many features of living circuits as we are aware, doing so entirely through self-organization.  Justification with respect to biology for each of the mechanisms can be found in their respective subsections within the \nameref{sec:Methods} section, however since reproducing cortical features and doing so through mechanisms were higher level priorities than always possessing biological analogs, some mechanisms and parameters merely reflect a particular design toward achieving those higher priority goals. Indeed since it is the case that certain biological mechanisms with respect to neuroplasticity are not fully understood and other attempts using only mechanisms with direct analogs required hand tuning \cite{miner2016plasticity}, the insertion of mechanisms with no direct known biological analog was necessary from the outset. In principle the mechanism for regulating firing rates from \cite{sweeney2015diffusive} may have been a candidate for the same role as MHP in our model since it resulted in heavy-tailed distributions of firing rates in simulated populations and possessed a plausible biological justification. However as of yet no direct experimental evidence has surfaced explicitly validating the model. Given the state of the field it was our goal to create a phenomenological mechanism not tied specifically to any specific biological counterpart. By formulating MHP in more abstract mathematical terms and ensuring that the collective results of the model were in line with known features of biological circuits, it was our aim that MHP might describe any of a diverse number of possible biological mechanisms, thus being complimentary to work along the lines of \cite{sweeney2015diffusive}.

%Furthermore, it must be understood that MANA does not necessarily represent the only or best way of achieving the results presented here, nor is it the goal here to decidedly prove such a thing is the case. The contribution of this work resides in it being the first such comprehensive model which attains the aforementioned goals, a proof of concept that such a thing is possible, and an example of how similar models moving forward might be approached and designed. We propose (with substantial evidence) that the equations laid out here make a good model of generic cortical circuits given the wide array of replicated phenomena and the fact that such phenomena were replicated from a null state. 
	 
Empirical studies aimed at finding certain key hallmarks of the particular formalisms here would be required to make more comprehensive statements as to MANA's validity as a scientific model, though. That is to say that our phenomenological aims do not entail that experimental evidence is unimportant. For instance the specific rule here for meta-homeostatic plasticity comprises a set of unambiguous formalisms translating the relationship between a neuron and its pre-synaptic neighbors into a realistic distribution of firing rates. The validity of those formalisms would require an empirical study to determine the relationships between the firing rates of pre- and post-synaptic neurons in living tissue, specifically looking for differences in mean firing rates that would match with those predicted by the equations. Such an experiment has not been carried out as of yet, and constitutes an avenue of future research. That being said, studies have been conducted on the synaptic properties of high firing rate neurons specifically, which MANA and MHP in particular appear to account for exceedingly well (See. Sec. \nameref{Sec.MANA_FR_wire} and Sub. Sec. \nameref{SubSec.MHP_FR_wire}).  The topology manifested by MHP specifically is consistent with and accounts for the peculiar features of wiring found by \cite{yassin2010embedded} and \cite{benedetti2012differential}  to be associated with high firing rate excitatory neurons.  While not absolute, this presents rather compelling evidence that MHP provides an accurate formalization of a (or many) biological mechanism(s) at play in determining average firing frequency in living circuits. 

In this way the elements of MANA are guides to what \emph{sorts of} mechanisms and relationships we might expect to find in living tissue. They are a conceptual scaffolding upon which network self-organization can be discussed and experiments formulated. The specific equations represent a hypothesis  for what the myriad of more complex lower-level biological mechanisms might be implementing or work in service of, which is supported by evidence in the form of its stability and broad range of replicated phenomena, but not grounded in experiment. But to be clear, a different phenomena being found to be at work in some part of a network can only fully invalidate its counterpart in MANA if it can be shown that such a phenomena is not in service of or performing the exact same function as its theoretical counterpart or offers greater explanatory power. Fundamentally MANA is a theoretical construct, and that nature defines its scope and capabilities. A good example of this lies in the fact that homeostatic modifications to synapses and firing thresholds generally take on the order of hours to days of real world time, whereas in MANA these changes occur on the order of seconds to minutes. This is a problem endemic to virtually all self-organizing models in this class \cite{lazar2009sorn, miner2016plasticity, litwin2014formation} and as pointed out in \cite{zenke2017hebbian} the model ``equivalents'' to these real world mechanisms should be called ``rapid compensatory processes' (RCPs)' for that reason. However in that very same review it is argued that RCPs are fundamentally necessary to stabilize Hebbian plasticity and must exist in some form alongside their slow counterparts in the real world. The implication being that some set of as-yet-discovered lower level mechanisms related to or interacting with the slower homeostatic processes must \emph{implementing} RCPs in one form or another. MANA can be thought of as a collection of just such RCPs, though it remains to be seen if MANA might be stable with simply much smaller learning rates for its processes. It is notable that in Fig. \ref{Fig3} after an initial rapid set of changes in thresholds and TFRs from the null state fluctuations in threshold tend to be very small indicating perhaps that once MHP organizes the distribution of firing rates and in conjunction with other processes orchestrates the firing rates of cells impinging on each neuron that rapid homeostatic plasticity may not be necessary for overall maintenance of activity within some acceptable range. This represents another avenue of future research for MANA. 
	 
\subsection*{Future Work}
	 
MANA opens up numerous avenues of future research. In fact this particular aspect of MANA can be considered to be its single strongest contribution. The original goal was to reverse engineer a neural circuit, which required that A) a wide variety of circuit features could be accounted for, and B) that those features had to be obtained via mechanism. The motivation for such an approach was to maximize the likelihood that whatever benefits such features and mechanisms conferred in biological circuits would be present in the artificial version. Having achieved (A) and (B), the next avenue of research clearly centers around ascertaining the computational properties of MANA, its applicability to different tasks, and the contributions of the different mechanisms and features to these. Indeed unpublished, preliminary results indicate that MANA can perform pattern separation on complex spatio-temporal inputs, and in particular action recognition tasks from video. 
	 
In addition to this more ambitious goal, a large amount of work remains with respect to exploring different aspects of MANA's quite large parameter space. While broad spectrum parameter searches are intractable here, smaller scale manipulations could provide valuable insight into the effects of the manipulations in the context of all of MANA's mechanisms. It has been demonstrated that, for instance, STDP becomes indistinguishable from a firing rate based mechanism when the pre- and post-synaptic neurons fire in a \emph{realistic} manner\cite{graupner2016natural}. A model possessing such a broad scope of mechanisms, which is known to produce realistic cortical features is ideal for providing a realistic setting for studying the effects of certain manipulations. 
	 
Models of this type, allow us to study how different inputs to the network may affect development, and indeed MANA can also be used as a tool for scientific modeling. MANA as presented here developed around a very artificial input stream meant only to provide some vague degree of statistical structure. However, this is far from the much richer set of inputs (and outputs) presented to real developing brain. By no means is our model an ``end all be all'', and indeed when given richer inputs, may reveal things still missing from our understanding. Von Melchner and colleagues rewired the retinal outputs of neonatal ferrets so as to ennervate their auditory rather than visual cortex, and found that not only did the ferrets' auditory cortices develop retinal maps normally found in visual cortex, but they responded to visual stimuli in ways consistent with visual, not auditory perception \cite{von2000visual}. This points to a certain genericness of cortex which can adapt to interpret arbitrary stimuli. Thus if our network model is "correct" then we should see it develop differently with different stimuli, and in ways which mirror the associated cortices in mammalian brains. Such avenues of research offer a "win-win" in that we either show our model to be correct or (hopefully) learn something about why it's incorrect. In the latter case self-organizing models give us a platform to test hypotheses. A model which self-organizes realistic cortical structure and behavior in a general sense is a template from which the mechanisms behind the specific structure and function of various brain regions can be studied. 	

\section*{Conclusion}

Based upon the results we can conclude that the formalisms laid out here which constitute MANA are indeed sufficient for the reproduction of a wide variety of known, highly complex  and nonrandom features of cortical circuits and thus the stated goals have been achieved. We have presented here a self-organizing model which builds upon prior work in the field by introducing a metaplastic mechanism which guides the self-organization of multiple plastic mechanisms in the network. We have also included inhibitory plasticity pervasively including inhibition of inhibition in a manner as yet seen in models of this type. The metaplastic architecture developed here including several regulatory mechanisms is both stable and capable of reproducing a wide variety of known features of living neural circuits.  While many topological features reproduced here can and have been explained by models like the SORN \cite{miner2016plasticity} these models have been unable to provide a phenomenological account of the development of TFRs which can account for the the features of synaptic topology and other properties known to be associated with highly active neurons \cite{benedetti2012differential}\cite{yassin2010embedded}. This is not limited to the particulars of properties known to be associated with neurons commensurate with their firing rates, and extends more generally into a phenomenological account of differentiation. Neurons which emerged from the MANA model not only possessed distinct qualities having to do with firing rate, but also had unique relationships to other cells in the network particularly between recurrent and input signals, implying distinct functional roles. Notably all of these features emerged entirely from \emph{mechanisms}. The only hand-tuning which could be said to have occurred with respect to the overall outcome and the network features of interest were parameters of mechanisms which produced those outcomes. In this way MANA represents the first complete phenomenological model of generic cortical circuitry in that it provides a set of functions which collectively take as input model neurons and synapses and produce as output an artificial cortical circuit which possesses a wide variety of features known to exist in living circuits.

\section*{Acknowledgments}
The authors thank Jeffrey Yoshimi and Ehren Newman for their comments on the manuscript and Jeffrey Yoshimi for creating Simbrain, which was used to develop the model and run experiments. This research was supported in part by the NSF Integrative Graduate Education and Research Traineeship (IGERT), Award Number: 0903495, NSF Robust Intelligence in Neurological Systems grant, Award Number: 1513779, and Lilly Endowment, Inc., through its support for the Indiana University Pervasive Technology Institute, and in part by the Indiana METACyt Initiative. The Indiana METACyt Initiative at IU is also supported in part by Lilly Endowment, Inc.

%\nolinenumbers

% Either type in your references using
% \begin{thebibliography}{}
% Text
% \end{thebibliography}
%
% or
%
% Compile your BiBTeX database using our plos2015.bst
% style file and paste the contents of your .bbl file
% here.
% 
%\bibliography{MANAbib}

\end{document}